\pdfoutput=1
\documentclass[a4paper,11pt,aps,prd,superscriptaddress,nofootinbib,floatfix]{revtex4-1}
\usepackage{microtype}
\usepackage{graphicx}
\graphicspath{{./figures/}}
\usepackage{subfigure}
\usepackage{booktabs}
\usepackage{tabularx}
\usepackage{comment}

\usepackage{mathtools}
\usepackage{amssymb}

\usepackage[utf8]{inputenc}
\usepackage[T1]{fontenc}
\usepackage[pdfusetitle,bookmarksopen,colorlinks,allcolors=blue]{hyperref}

\usepackage{xcolor}

\begin{document}
\title{Fourier analysis of  compact oscillons}

\author{P.~Klimas}
\email{pawel.klimas@ufsc.br}
\affiliation{Departamento de Física, Universidade Federal de Santa Catarina,\\
Campus Trindade, 88040-900, Florianópolis, Brazil}

\author{L. R. Livramento~}
\email{livramento@usp.br}
\affiliation{Instituto de F\'\i sica de S\~ao Carlos,\\
Universidade de S\~ao Paulo, 13560-970, S\~ao Carlos-SP, Brazil\\}
\affiliation{Institute of Theoretical Physics,  Jagiellonian University,
{\L}ojasiewicza 11, Krak\'{o}w, Poland}

\author{J.~S.~Streibel}
\email{jstreibel@gmail.com}
\affiliation{Departamento de Física, Universidade Federal de Santa Catarina,\\
Campus Trindade, 88040-900, Florianópolis, Brazil}

\begin{abstract}
We present a new derivation of the exact compact oscillon in the signum-Gordon model using Fourier methods. This approach decomposes the oscillon into an infinite, uniformly convergent series of discrete modes consisting of a fundamental frequency and its odd higher harmonics, in contrast to standard analytic potentials, where the corresponding infinite perturbative series fails to converge. Crucially, our oscillon evades the non-existence theorem for spatially localized, small-amplitude periodic states. We investigate the accuracy of truncating the Fourier series and find that the primary mode alone provides a remarkably accurate approximation of the exact solution, with agreement improving significantly when the first two non-vanishing harmonic modes are included. Furthermore, we generalize this framework to a swaying compact oscillon with a periodically moving boundary. We show that describing such an oscillon requires even-order modes, whose coupling with the odd source terms induces a dynamic modulation that generates the swaying motion, even though the spatial Fourier profiles themselves are time-independent.

\end{abstract}

\maketitle
\tableofcontents

\section{Introduction}

Oscillons are spatially localized, long-lived, and periodic finite-energy field configurations found across a wide variety of scalar field models. Originally discovered in a Higgs-type scalar field model by Bogolyubsky and Makhankov, these spherically symmetric configurations were initially called "pulsons" \cite{Bogolyubsky:1976nx, Bogolyubsky:1976yu}. The terminology later evolved, and the name "oscillon" was widely popularized by Gleiser and his collaborators \cite{Gleiser:2009ys}.

One of the most striking properties associated with these configurations is their remarkably long lifetime. Such non-decaying or long-lived behavior is inherently expected in integrable field theories, where the existence of exact, infinitely surviving breathers \cite{PhysRevLett.42.937, Ivancevic2013, doi:10.1142/S0217751X95000917} is rigorously guaranteed by an infinite number of conserved charges originating from hidden symmetries. Remarkably, these periodic, localized structures survive even when the system is no longer strictly integrable, as shown in quasi-integrable field models \cite{Ferreira_2016}.

On the other hand, the emergence of long-lived breathers in non-integrable scalar field theories -- such as the standard $\phi^4$ model -- presents a far more intriguing puzzle. Unlike topological solitons (which are protected by boundary conditions) or non-topological solitons (which are stabilized by a conserved Noether charge), oscillons possess no obvious conserved quantity to guarantee their extraordinary stability. Consequently, understanding the dynamical mechanisms that allow these localized configurations to resist radiation damping and maintain their coherence over vast timescales remains a central and captivating question in non-linear field dynamics. In this context, the analysis of normal modes of small-amplitude oscillons was recently investigated in \cite{Evslin:2024sup}.

One remarkable approach to this subject is presented by Fodor and collaborators, who describe oscillons as weakly localized quasi-breathers \cite{PhysRevD.78.025003}. In \cite{PhysRevD.74.124003}, the authors propose a small-amplitude expansion as a more systematic approach to describing oscillons in scalar field theories. These long-lived, breather-like objects are remarkably well approximated by asymptotic series. Furthermore, the analytically designed initial data based on this approach give rise to numerically evolved, extremely long-lived oscillons.

The appearance of oscillons in scalar field theories arises in several distinct contexts. One pathway involves properly chosen initial data, as discussed in \cite{PhysRevD.74.124003}. However, oscillons are also frequently generated via dynamical processes, such as kink--antikink collisions \cite{Romanczukiewicz:2010eg, BAZEIA2020135291, Hahne:2024qby}. Furthermore, recent results have demonstrated a link between oscillons and certain non-topological localized objects known as Q-balls through a renormalization group scheme \cite{PhysRevLett.134.081601}. 

Ultimately, the precise mechanism driving the existence of oscillons and their outstanding stability remains an open question. Several studies suggest the crucial role of a mass threshold within the small-amplitude perturbative framework, alongside the requirement that the self-interaction potential must be flatter than its quadratic counterpart near the vacuum. While these conditions have been thoroughly established for standard field theories featuring smooth, analytic potentials at the minimum, their validity is not necessarily universal. This limitation becomes especially prominent in alternative field-theoretic frameworks, such as models governed by V-shaped, non-analytic potentials \cite{Arodz:2002yt} -- a canonical example of which is the signum-Gordon model. 

The exact oscillon solution in the signum-Gordon model was discovered in \cite{Arodz:2007jh}. It is a compact structure \cite{PhysRevLett.70.564} of strictly finite size that is perfectly periodic and features an infinite lifetime. When unperturbed, it does not emit any radiation -- a remarkable departure from standard oscillons, which are only quasi-stable and inevitably suffer from small, continuous radiation leakage.
The signum-Gordon potential, given by $V(\phi) = \vert{}\phi\vert{}$, features a sharp, stable minimum at $\phi = 0$. Furthermore, the model possesses an exact scaling symmetry. Under the universal scaling relation
\begin{equation}\label{eq:scalling}
\phi^{(\lambda)}(t,x) = \lambda^2 \phi\left(\frac{t}{\lambda}, \frac{x}{\lambda}\right),
\end{equation}
any existing oscillon solution can be scaled to generate a continuous family of oscillons spanning arbitrary amplitudes (both small and large) and energy scales. Another striking difference is that the signum-Gordon potential is not flatter than its quadratic counterpart; on the contrary, it is significantly sharper near the minimum. Consequently, the traditional mass threshold -- typically formulated through the perturbative mass $m^2 = \left.\frac{d^2V}{d\phi^2}\right\vert{}_{\phi=0}$ -- is not well-defined here, though it can be formally treated as infinite within certain regularization schemes. 

When perturbed, compact oscillons do radiate; however, this radiation is of a fundamentally different nature than the radiation fields typically emitted in models with smooth potentials. Rather than losing energy through a continuum of linear dispersive waves, compact oscillons release their excess energy by emitting smaller, discrete compact oscillons. These emitted structures can, in turn, interact with one another to produce further compact configurations.

This characteristic cascading process has been observed both in the direct scattering of compact oscillons \cite{Hahne:2019ela} and during the decay of shock waves within the signum-Gordon model \cite{Hahne:2019odw}. Similarly, the scattering of compact kinks in models that support topological charges has been shown to efficiently generate these localized states \cite{Hahne:2023dic, Hahne:2022wyl}. Collectively, all available results concerning compact oscillons strongly suggest that they play a foundational role in the non-linear dynamics of the field, acting as the primary vacuum excitations of the signum-Gordon model and related field theories of the same class. Because the broken analyticity at the minimum prevents the formation of standard plane-wave radiation, these compact pulson configurations replace linear waves as the natural, localized degrees of freedom through which the system redistributes energy and approaches equilibrium. 

Despite the superficial simplicity of the signum-Gordon equation, the analytical treatment of its solutions is remarkably challenging in practice. The core difficulty stems from the model's peculiar non-linearity, which strictly forbids linearization in the limit of small field amplitudes. Consequently, even small-amplitude excitations remain fundamentally non-linear. This feature becomes particularly acute when considering the scattering of two compact objects, as the principle of linear superposition fails completely regardless of how small the field amplitude is.
Another major technical hurdle arises because solutions -- such as the exact compact oscillon -- are constructed by matching multiple distinct algebraic patches (so-called "partial solutions") whose spatial boundaries are non-trivial functions of time. Tracking these moving boundaries introduces significant analytical inconveniences when working with compactons in non-analytic potentials. In this paper, we overcome these long-standing technical difficulties by proposing an alternative approach to signum-Gordon oscillons based on Fourier methods.

The paper is organized as follows. In Section \ref{sec:simplest_oscillon}, we present the case of the simplest compact oscillon with static boundaries and analyze its corresponding energy and momentum densities. In Section \ref{sec:fourierbasic}, we develop the formal derivation of the compact oscillon by employing a Fourier series ansatz, which decomposes the patched partial solutions into an infinite set of interacting modes. In this section, we also investigate the successive approximations of the oscillon using a finite number of modes. Within this framework, we analyze the hypothesis of whether the compact oscillon can be understood as a composite object -- specifically, a bound state of two localized structures. Finally, in Section \ref{sec:Fourierswaying}, we study the generalization of the Fourier decomposition to an oscillon with moving boundaries. In Section \ref{sec:summary}, we summarize our findings and draw our conclusions.

\section{Compact exact oscillon in the signum-Gordon model}
\label{sec:simplest_oscillon}
\subsection{The basic compact oscillon}

The simplest compact oscillon solution was initially reported in \cite{Arodz:2007jh}, where it was derived as the solution to a specific Cauchy problem given a properly selected set of initial conditions, $\{\phi(0,x), \partial_t\phi(0,x)\}$. The configuration is periodic and constructed by appropriately matching a set of quadratic polynomials defined over compact domains in (1+1)-dimensional Minkowski spacetime. Due to the intrinsic periodicity of the oscillon, it is sufficient to define the constituent components within a single fundamental period, $T=1$. Letting $\phi_0(t,x)$ denote the solution during this fundamental period, the subsequent temporal segments $\phi_k(t,x)$ of the oscillon for $t \ge T$ are generated via the translation $\phi_k(t,x) = \phi_0(t - kT, x)$ for $kT \le t \le (k+1)T$, where $k \in \mathbb{Z}$. Oscillons with arbitrary periods $T=\lambda$ can then be obtained through the scaling transformation \eqref{eq:scalling}.

The signum-Gordon equation in $(1+1)$ dimensions is given by
\begin{equation}
\partial_t^2 \phi - \partial_x^2 \phi + \text{sgn}(\phi) = 0.
\end{equation}
To exploit the spatial parity symmetry of the configuration about the core of the oscillon, we define the solution over the centralized, symmetric spatial segment $-T/2 \le x \le T/2$ for the duration of the fundamental period $0 \le t \le T$. Taking advantage of the exact spatial reflection symmetry $\phi(t,-x) = \phi(t,x)$, the profile is constructed from a seed function $\varphi(t, x)$ restricted to the first quarter-period, $0 \le t \le T/4$. By introducing the absolute value $|x|$, the five original piecewise regions collapse elegantly into three symmetric, compact patches:

\begin{equation}\label{compact_patches}
\varphi(t,x) = \begin{cases} 
-t^2 + \frac{T}{2}t - \frac{x^2}{2} & \text{if } |x| \le t \\[1.5ex]
t\left(\frac{T}{2} - |x|\right) - \frac{t^2}{2} & \text{if } t < |x| \le \frac{T}{2} - t \\[1.5ex]
\frac{1}{2}\left(\frac{T}{2} - |x|\right)^2 & \text{if } \frac{T}{2} - t < |x| \le \frac{T}{2} 
\end{cases}
\end{equation}

The full spatiotemporal dynamics of the exact compact oscillon $\phi(t,x)$ over the complete fundamental period are then established by extending this core seed profile via discrete time-translations and sign inversions across successive quarter-periods:

\begin{equation}\label{global_solution}
\phi(t,x) = \begin{cases}
\varphi(t, x) & \text{if } 0 \le t \le \frac{T}{4} \\[1.5ex]
\varphi\left(\frac{T}{2} - t, x\right) & \text{if } \frac{T}{4} < t \le \frac{T}{2} \\[1.5ex]
-\varphi\left(t - \frac{T}{2}, x\right) & \text{if } \frac{T}{2} < t \le \frac{3T}{4} \\[1.5ex]
-\varphi\left(T - t, x\right) & \text{if } \frac{3T}{4} < t \le T
\end{cases}
\end{equation}

This formulation guarantees that the field profile and its first derivatives remain perfectly continuous across all internal moving boundaries, thereby consistently satisfying the field equation throughout the compact domain.

\subsection{Energy and Momentum Density}
\label{sec:densities}

The Signum-Gordon model has the Lagrangian density
\begin{equation}
{\cal L} = \frac{1}{2}(\partial_t\phi)^2 - \frac{1}{2}(\partial_x\phi)^2 - |\phi|,
\end{equation}
which gives the energy-momentum tensor for the scalar field $\phi(t,x)$ via the standard definition
\begin{equation}
T^{\mu\nu} = \partial^\mu\phi \partial^\nu\phi - \eta^{\mu\nu}{\cal L},
\end{equation}
where $\eta^{\mu\nu} = \text{diag}(1, -1)$ is the Minkowski metric tensor. From the components of $T^{\mu\nu}$, we get the energy density $\mathcal{T}^{00} \equiv \mathcal{H}$ as
\begin{equation}
\mathcal{H} = \frac{1}{2}(\partial_t\phi)^2 + \frac{1}{2}(\partial_x\phi)^2 + |\phi|,
\end{equation}
and the linear momentum density $\mathcal{T}^{01} \equiv \mathcal{P}$ as
\begin{equation}\label{eq:momdensity}
\mathcal{P} = -\partial_t\phi \partial_x\phi.
\end{equation}

By utilizing the spatial reflection parity of the field profile, the continuous energy density for the seed configuration $\mathcal{H}_{\text{s}}(t,x)$ during the first quarter-period $0 \le t \le T/4$ can be written in a compact three-patch absolute-value form:
\begin{equation}\label{eq:enseed_final}
\mathcal{H}_{\text{s}}(t,x) = \begin{cases} 
t^2 - \frac{T}{2}t + \frac{T^2}{8} & \text{if } |x| \le t \\[1.5ex]
\frac{1}{2}\left(\frac{T}{2} - |x|\right)^2 + \frac{1}{2}t^2 & \text{if } t < |x| \le \frac{T}{2} - t \\[1.5ex]
\left(\frac{T}{2} - |x|\right)^2 & \text{if } \frac{T}{2} - t < |x| \le \frac{T}{2} 
\end{cases}
\end{equation}

The full spatiotemporal evolution of the energy density $\mathcal{H}(t,x)$ over the entire fundamental period $0 \le t \le T$ is then achieved by matching the transformations of the underlying fields across successive quarter-periods:
\begin{equation}\label{eq:global_energy_final}
\mathcal{H}(t,x) = \begin{cases}
\mathcal{H}_{\text{s}}(t, x) & \text{if } 0 \le t \le \frac{T}{4} \\[1.5ex]
\mathcal{H}_{\text{s}}\left(\frac{T}{2} - t, x\right) & \text{if } \frac{T}{4} < t \le \frac{T}{2} \\[1.5ex]
\mathcal{H}_{\text{s}}\left(t - \frac{T}{2}, x\right) & \text{if } \frac{T}{2} < t \le \frac{3T}{4} \\[1.5ex]
\mathcal{H}_{\text{s}}\left(T - t, x\right) & \text{if } \frac{3T}{4} < t \le T
\end{cases}
\end{equation}


Correspondingly, the linear momentum density for the seed configuration $\mathcal{P}_{\text{s}}(t,x)$ in this compact absolute-value notation takes the form:
\begin{equation}\label{eq:momseed_final}
\mathcal{P}_{\text{s}}(t,x) = \begin{cases} 
-x \left(2t - \frac{T}{2}\right) & \text{if } |x| \le t \\[1.5ex]
t \left(\frac{T}{2} - |x| - t\right) \text{sgn}(x) & \text{if } t < |x| \le \frac{T}{2} - t \\[1.5ex]
0 & \text{if } \frac{T}{2} - t < |x| \le \frac{T}{2} 
\end{cases}
\end{equation}

The full spatiotemporal evolution of the linear momentum density $\mathcal{P}(t,x)$ over the entire fundamental period $0 \le t \le T$ is then achieved by matching the transformations of the underlying partial derivatives across successive quarter-periods:
\begin{equation}\label{eq:global_momentum_final}
\mathcal{P}(t,x) = \begin{cases}
\mathcal{P}_{\text{s}}(t, x) & \text{if } 0 \le t \le \frac{T}{4} \\[1.5ex]
-\mathcal{P}_{\text{s}}\left(\frac{T}{2} - t, x\right) & \text{if } \frac{T}{4} < t \le \frac{T}{2} \\[1.5ex]
\mathcal{P}_{\text{s}}\left(t - \frac{T}{2}, x\right) & \text{if } \frac{T}{2} < t \le \frac{3T}{4} \\[1.5ex]
-\mathcal{P}_{\text{s}}\left(T - t, x\right) & \text{if } \frac{3T}{4} < t \le T
\end{cases}
\end{equation}

\begin{figure}[h!]
\subfigure[]{\includegraphics[width=0.31\linewidth]{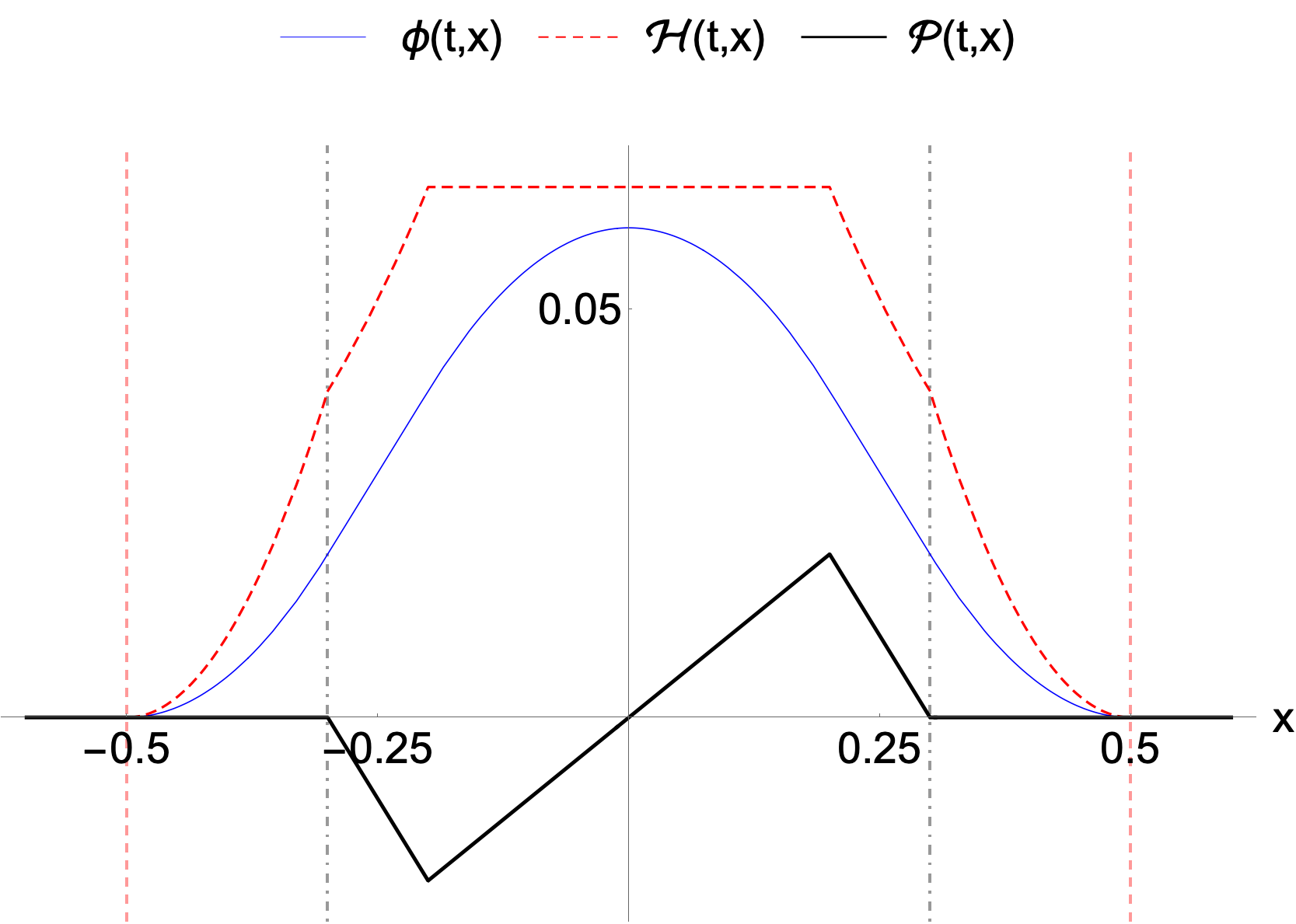}\label{fig:020}}
\subfigure[]{\includegraphics[width=0.31\linewidth]{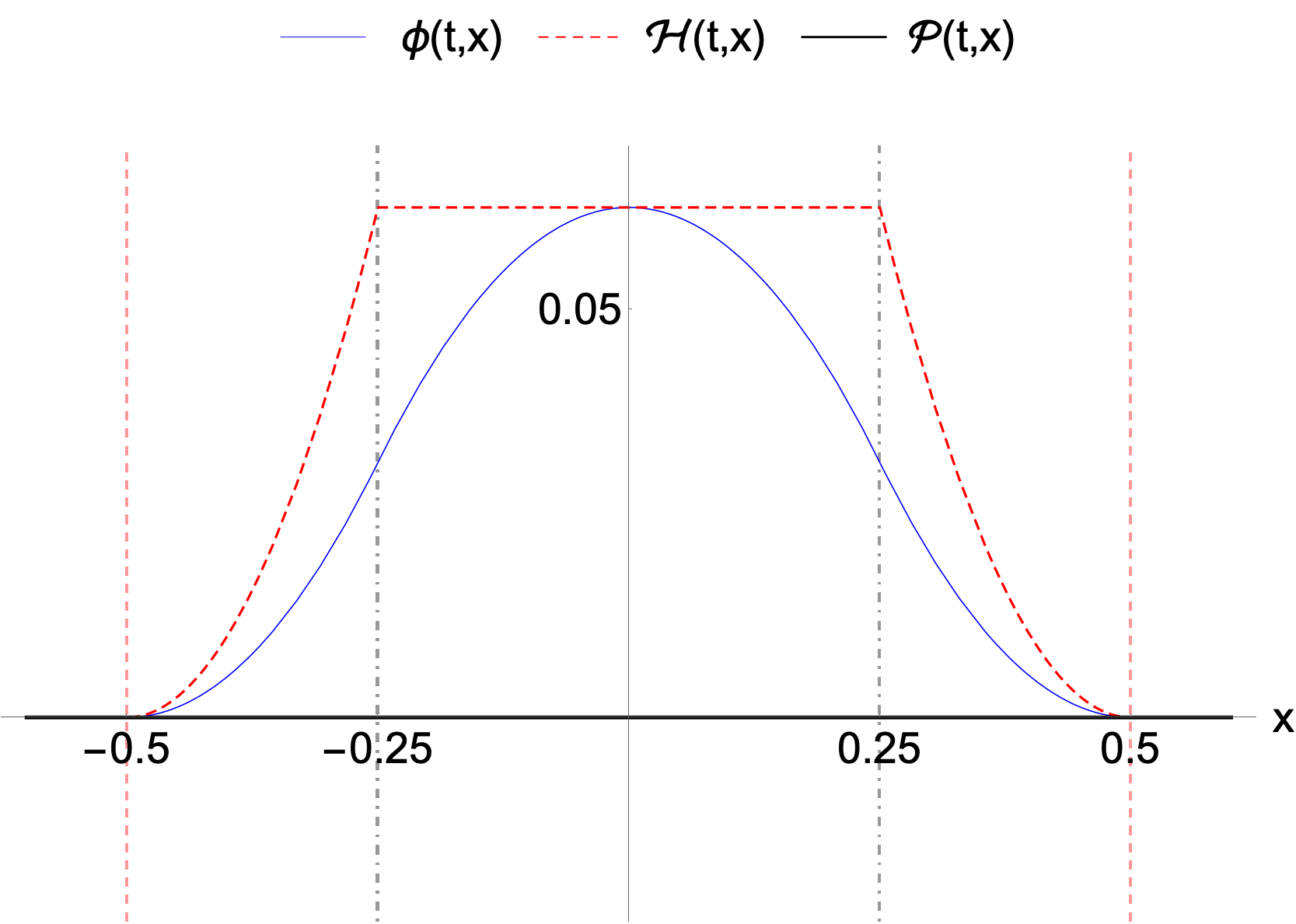}\label{fig:025}}
\subfigure[]{\includegraphics[width=0.31\linewidth]{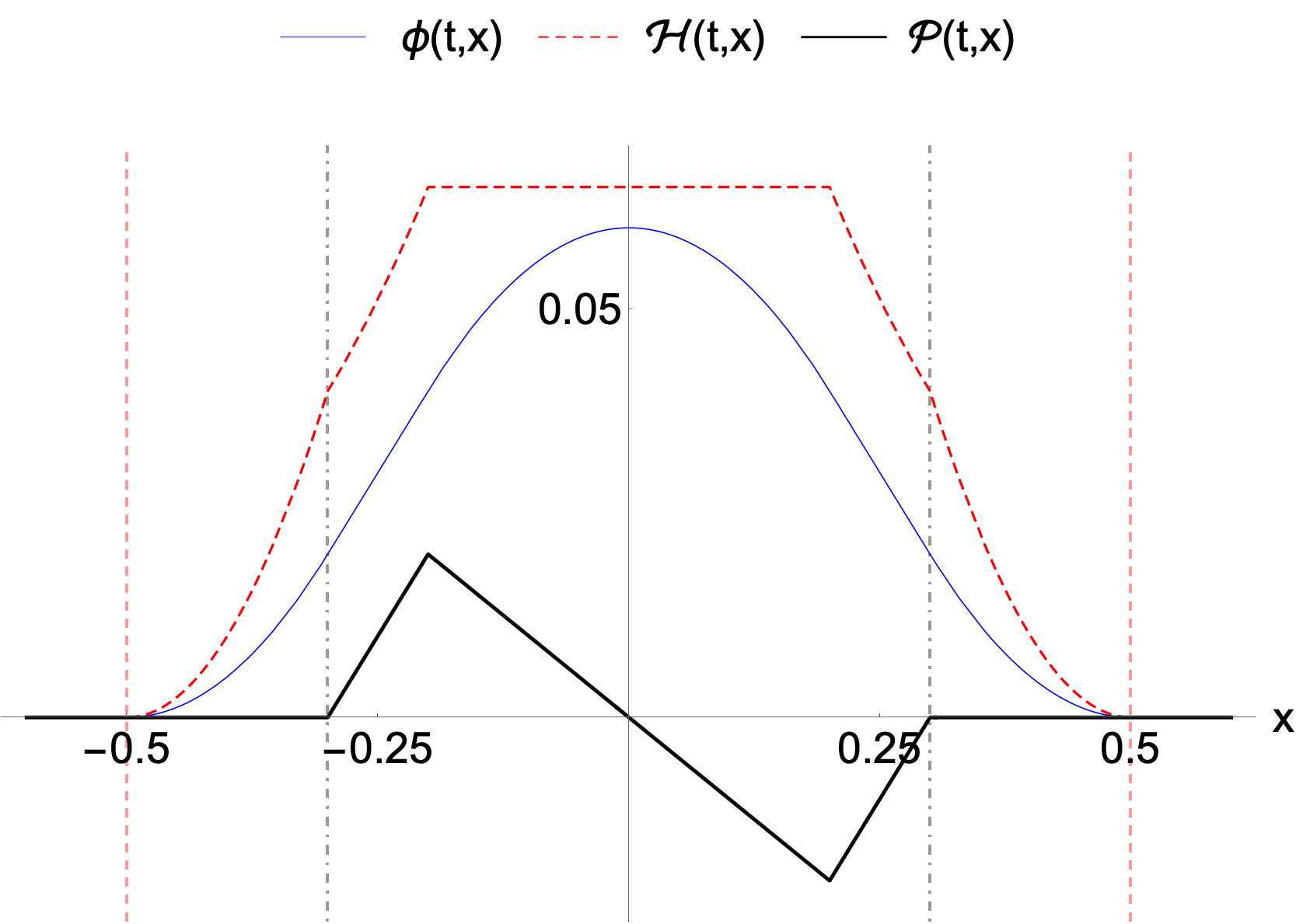}\label{fig:030}}
\caption{Snapshots of the exact oscillon field $\phi$, its energy density $\mathcal{H}$, and linear momentum density $\mathcal{P}$ at (a) $t=0.2$, (b) $t=0.25$, and (c) $t=0.3$. Here $T=1$.}\label{fig:densities}
\end{figure}

The momentum density vanishes everywhere at times $t_j = j\frac{T}{4}$ (for $j = 0, 1, 2, \dots$), which correspond exactly to the instances when either the time derivative or the spatial derivative of the field vanishes globally. Specifically, at the even quarter-periods ($t = 0$ and $t = T/2$), the field profile passes through zero, which implies that its spatial gradient vanishes, $\partial_x\phi = 0$. At the odd quarter-periods ($t = T/4$ and $t = 3T/4$), the field reaches its maximal spatial amplitude and is instantaneously at rest, $\partial_t\phi = 0$. Since the linear momentum density is given by the canonical relation \eqref{eq:momdensity}, the fact that one of these two derivatives is uniformly zero at each $t_j$ ensures that the momentum density profile vanishes entirely at these specific quarter-period intervals.

Notably, the partial solutions that form the most external part of the oscillon are determined by $\pm\frac{1}{2}\left(\frac{T}{2} - |x|\right)^2$, which have the form of static parabolas. Consequently, in the regions determined by these solutions, the momentum density vanishes identically. This implies that the momentum density support is a geometric region limited by $|x| \le z(t)$. To formalize this boundary, we first define the symmetric zig-zag profile on the fundamental half-period interval $t \in [0, T/2]$ as:
\begin{equation}\label{eq:support_boundary}
z(t) = \frac{T}{4} + \left| t - \frac{T}{4} \right|.
\end{equation}
Since the system's dynamics are periodic, this $z(t)$ segment defined on $[0, T/2]$ repeats periodically in each subsequent half-period interval.

The spatial antisymmetry of the linear momentum density profile, $\mathcal{P}(t,-x) = -\mathcal{P}(t,x)$, justifies treating the localized compact oscillon -- at least hypothetically at this stage -- as a composite system of two interacting substructures: one localized on the left ($x < 0$) and the other on the right ($x > 0$).

To characterize the internal motion  of these subsystems, we define the collective center-of-mass coordinates for the left ($X_L$) and right ($X_R$) sub-objects as the normalized first spatial moments of the energy density $\mathcal{H}(t,x)$:
\begin{equation}\label{eq:cm_definitions}
X_L(t) = \frac{1}{E_L} \int_{-T/2}^{0} x \,\mathcal{H}(t,x) \, dx, \qquad X_R(t) = \frac{1}{E_R} \int_{0}^{T/2} x \,\mathcal{H}(t,x) \, dx,
\end{equation}
where $E_L$ and $E_R$ denote the sub-object masses (energies) contained within their respective half-domains. Owing to the exact spatial reflection symmetry of the global energy density profile $\mathcal{H}(t,-x) = \mathcal{H}(t,x)$, these subsystem energies are strictly conserved, time-independent constants satisfying:
\begin{equation}\label{eq:sub_energies}
E_L = E_R = \frac{E}{2} = \frac{T^3}{48}.
\end{equation}
Furthermore, the spatial parity yields a rigid constraint between the trajectories, forcing $X_L(t) = -X_R(t)$ for all $t$. 

By inserting the continuous piecewise configuration of the seed energy density $\mathcal{H}_{\text{s}}(t,x)$ into Eq.~\eqref{eq:cm_definitions} and performing the explicit spatial integrations across the three absolute-value patches, the explicit analytical trajectory of the right-hand center of mass during the first quarter-period $0 \le t \le T/4$ simplifies uniquely to:
\begin{equation}\label{eq:xr_seed_final}
X^{(s)}_R(t) = \frac{T}{8} + \frac{3}{T}t^2 - \frac{12}{T^2}t^3 + \frac{12}{T^3}t^4.
\end{equation}

The full periodic spatiotemporal evolution of the collective sub-object coordinates over the complete fundamental cycle $0 \le t \le T$ is then reconstructed via discrete time-reflections and shifts mapping the core quarter-period seed trajectory:
\begin{equation}\label{eq:xr_global_final}
X_R(t) = \begin{cases}
X^{(s)}_R(t) & \text{if } 0 \le t \le \frac{T}{4} \\[1.5ex]
X^{(s)}_R\left(\frac{T}{2}-t\right) & \text{if } \frac{T}{4}\le t \le \frac{T}{2}  \\[1.5ex]
X^{(s)}_R\left(t - \frac{T}{2}\right) & \text{if } \frac{T}{2} \le t \le\frac{3T}{4}  \\[1.5ex]
X^{(s)}_R\left(T-t\right) & \text{if } \frac{3T}{4} \le t \le T \\[1.5ex]
\end{cases}.
\end{equation}
\begin{figure}[h!]
\subfigure[]{\includegraphics[width=0.45\linewidth]{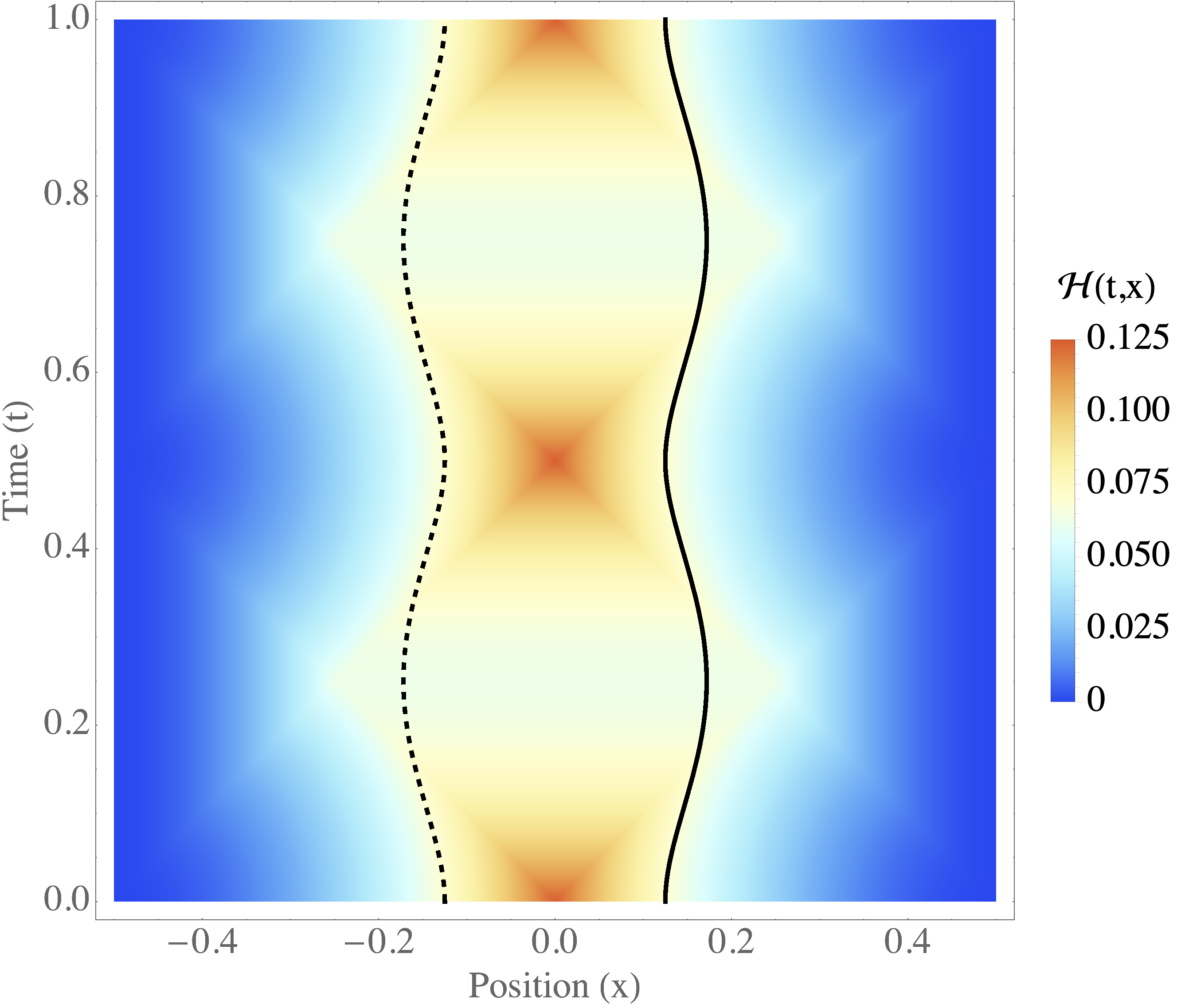}\label{fig:traj-energy}}\;\;
\subfigure[]{\includegraphics[width=0.45\linewidth]{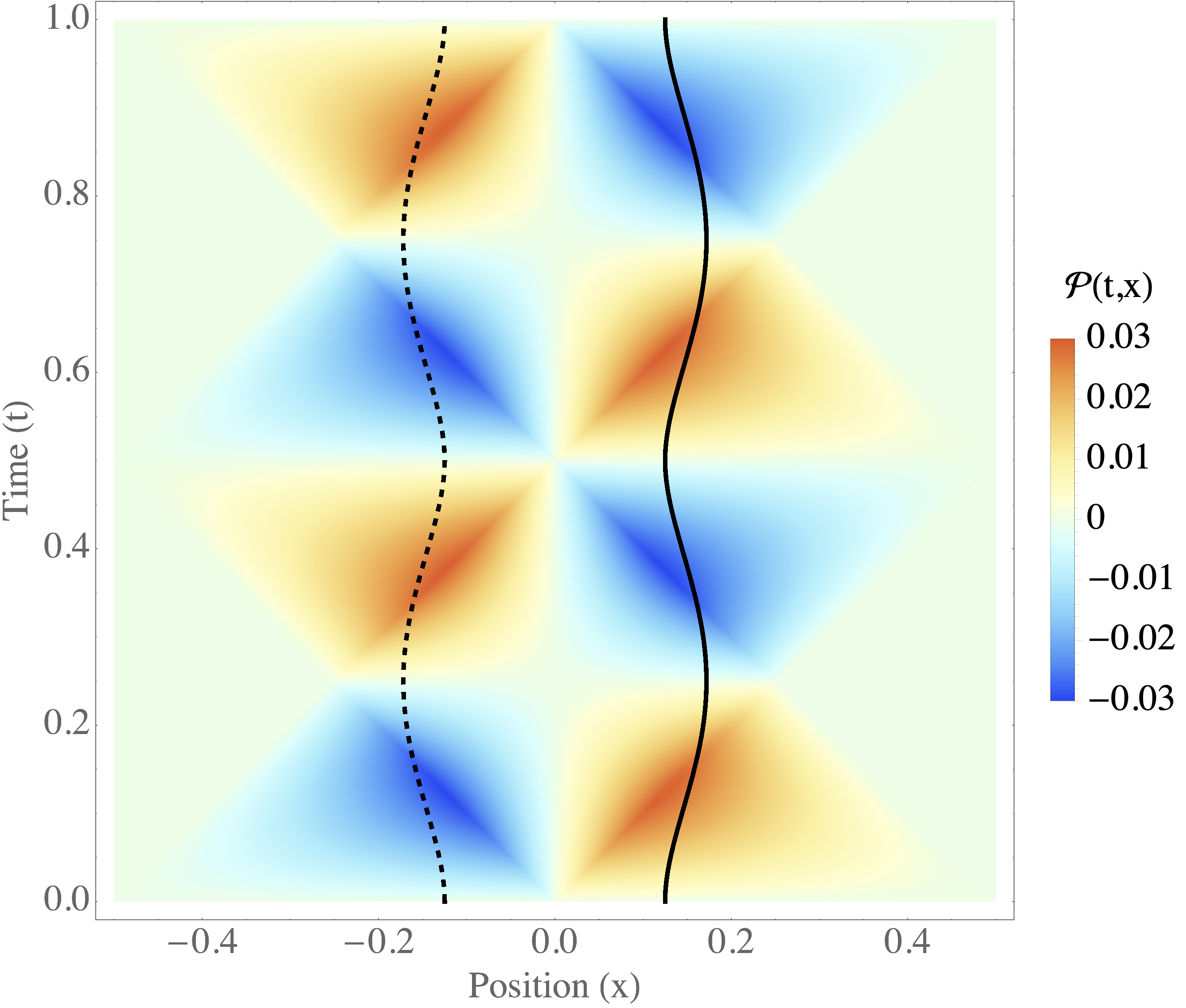}\label{fig:traj-momentum}}
\caption{Spatiotemporal trajectories of the centers of mass for the left (dashed line) and right (solid line) subsystem components overlaid on the background profiles of (a) the energy density $\mathcal{H}(t,x)$ and (b) the linear momentum density $\mathcal{P}(t,x)$.}
\end{figure}

Direct evaluation of the temporal derivative of Eq.~\eqref{eq:xr_seed_final} confirms that the collective velocity vanishes exactly at the quarter-period interfaces, i.e., $\dot{X}_R(\frac{n T}{4}) = 0$, $n\in N$. The resulting trajectories $X_{L,R}(t)$ periodic paths describing the continuous expansion and contraction cycles of the internal structure. The oscillation of the right-hand trajectory is localized around the central value $X_{0R} = \frac{19}{128}T$, with an amplitude given by $\Delta X_R = X_R\left(\frac{T}{4}\right) - X_R(0) = \frac{3}{64}T$. By symmetry, the left-hand trajectory satisfies $X_L(t) = -X_R(t)$.

The positive (negative) value of the momentum density for $x > 0$ ($x < 0$) means that the momentum is flowing outward in the direction of the external oscillon boundary, as visualized by the warm regions in Fig.~\ref{fig:traj-momentum}. Conversely, when the signs invert, the momentum is flowing inward in the direction of the oscillon center at $x = 0$. We can conclude, therefore, that the internal oscillon dynamics exhibit a much more complex behavior than a simple collective oscillation of field values. This behavior of the momentum density closely resembles a scenario featuring two bounded objects that alternately approach and recede from each other, undergoing a continuous cycle of compression and expansion, a feature tightly tracked by the collective paths shown in Fig.~\ref{fig:traj-energy} and ~\ref{fig:traj-momentum}.To determine whether this exact oscillon is a composite object, we examine its Fourier spectrum. This is addressed in the next section.


\section{Fourier analysis of the basic compact oscillon}
\label{sec:fourierbasic}
\subsection{Derivation of compact oscillon}

In this section, we present an alternative analytical framework for solving the signum-Gordon equation:\begin{equation}
\partial^2_t\phi - \partial^2_x\phi + \text{sgn}(\phi) = 0. 
\label{eq:sg}\end{equation}The proposed method relies on constructing a periodic, spatially compact solution, denoted as $\phi(t,x)$, by utilizing a formulation that combines Fourier series expansions with Fourier transform techniques. To formalize this framework, we impose the following mathematical and physical boundary conditions and symmetries on the scalar field $\phi(t,x)$:
\begin{enumerate}
\item \textbf{Spatial compact support:} The solution possesses compact support on the spatial interval $|x| \le \frac{L}{2}$, outside of which it vanishes identically:
\begin{equation}
\phi(t,x) = 0 \quad \text{for} \quad |x| > \frac{L}{2}.
\end{equation}

\item \textbf{Temporal periodicity:} The solution is periodic in time with a fundamental period $T$, satisfying:
\begin{equation}
\label{cond1}
\phi(t+T, x) = \phi(t,x),
\end{equation}
where the corresponding fundamental frequency is defined as $\omega \equiv \frac{2\pi}{T}$.

\item \textbf{Temporal nodes and symmetries:} The solution vanishes identically at the temporal nodes $t_j = \frac{jT}{2}$ for $j \in \mathbb{Z}$. It exhibits odd temporal parity, satisfying:
\begin{equation}
\label{cond2}
\phi(-t, x) = -\phi(t,x).
\end{equation}
Additionally, the simplest oscillon with static boundaries exhibits a half-period sign-reversal symmetry (anti-periodicity):
\begin{equation}
\label{cond3}
\phi\left(t + \frac{T}{2}, x\right) = -\phi(t,x).
\end{equation}
\end{enumerate}

Given its temporal periodicity, the solution can be represented generally as a complex Fourier series:
\begin{equation}\label{complex_fourier}\phi(t,x) = \sum_{n=-\infty}^{\infty} \widetilde{\phi}_{n}(x) e^{i \omega_n t},\qquad \text{where}\qquad \omega_n\equiv\frac{2\pi}{T} n.
\end{equation}
Imposing the odd parity constraint from \eqref{cond2} implies that $\widetilde{\phi}_0(x) = 0$ and $\widetilde{\phi}_{-n}(x) = -\widetilde{\phi}_n(x)$. Assuming $\phi(t,x)$ is a real-valued field, these relations yield $\widetilde{\phi}_n(x) = -\widetilde{\phi}_n^*(x)$, confirming that the coefficients are purely imaginary. This allows us to reduce \eqref{complex_fourier} to a real sine series within the domain of support:
\begin{equation}\label{series}
\phi(t,x) = \sum_{n=1}^{\infty} b_n(x) \sin( \omega_n t),
\end{equation}
where the real spatial profiles are defined via $b_n(x) \equiv 2i \widetilde{\phi}_{n}(x)$. Under the assumption that $\phi(t,x)$ does not change sign within the interval $t \in (0, T/2)$ for $|x| < \frac{L}{2}$, the signum term $\text{sgn}(\phi(t,x))$ inherently shares the same spatial support and temporal parity as the field itself. Consequently, it can be factored into a product of independent spatial and temporal modulation terms:
\begin{equation}
\text{sgn}\big(\phi(t,x)\big) = \Pi(x) \Sigma(t),
\end{equation}
where $\Pi(x)$ is a rectangular boxcar window function:
\begin{equation}
\label{window}
\Pi(x) := \theta\left(x + \frac{L}{2}\right) - \theta\left(x - \frac{L}{2}\right) = 
\begin{cases} 
1 & |x| \le \frac{L}{2} \\ 
0 & |x|>\frac{L}{2} 
\end{cases},
\end{equation}
with $\theta(x)$ denoting the Heaviside step function. The explicit inclusion of the window function $\Pi(x)$ represents a {\it compactness condition} imposed on the target solution.

The temporal component, $\Sigma(t) := \text{sgn}(\sin(\omega t))$, represents a periodic square wave with odd parity, which can be explicitly expressed via its Fourier series expansion:
\begin{equation}\label{sigma}
\Sigma(t) =\sum_{n=1,3,5,\ldots}^{\infty} \frac{4}{\pi n} \sin(\omega_n t).
\end{equation}

Because the solution satisfies the half-period sign-reversal symmetry defined in condition~\eqref{cond3}, its temporal Fourier expansion is restricted exclusively to odd harmonics ($n = 1, 3, 5, \dots$). This spectral structure is entirely consistent with the driving signum source $\Sigma(t) = \text{sgn}(\sin \omega t)$, which likewise contains only odd modes. Consequently, within this symmetry sector, the scalar field may be represented purely as a Fourier sine series over odd-order harmonics.

To exploit this spectral localization, we express the spatial profiles $b_n(x)$ and the window function $\Pi(x)$ via their inverse Fourier transforms:
\begin{equation}
b_n(x) = \frac{1}{2\pi} \int_{-\infty}^{\infty} \hat{b}_n(k) e^{ikx} \, dk, \quad \text{and} \quad \Pi(x) = \frac{1}{2\pi} \int_{-\infty}^{\infty} \hat{\Pi}(k) e^{ikx} \, dk,
\end{equation}
where 
\begin{equation}\label{eq:transformPi}
\hat{\Pi}(k) = \frac{2\sin(\frac{kL}{2})}{k}.
\end{equation}
Substituting these integral representations along with the Fourier series for $\phi(t,x)$ and $\text{sgn}(\phi)$ into the signum-Gordon equation \eqref{eq:sg} yields:
\begin{equation}\label{eq:tansformedequation}
\sum_{n=1,3,\ldots}^{\infty} \sin(\omega_n t) \int_{-\infty}^{\infty} \left[ \left(k^2 - n^2\omega^2\right)\hat{b}_n(k) + \frac{4}{\pi n}\hat{\Pi}(k) \right] e^{ikx} \, dk = 0.
\end{equation}
By invoking the orthogonality of the harmonic basis $\sin(\omega_n t)$ and the uniqueness of the Fourier transform, the signum-Gordon equation projectively transforms into a system of decoupled algebraic equations for each odd harmonic $n$ in the wavenumber $k$-space. Solving this system and explicitly substituting \eqref{eq:transformPi} yields the spectral components of the field profiles:
\begin{equation}
\hat{b}_n(k) = -\frac{8}{\pi n} \frac{\sin\left(\frac{kL}{2}\right)}{k\left(k^2 - \omega_n^2\right)},
\label{eq:phi_hat_solution}
\end{equation}
where $\omega_n = n\omega$. Expression \eqref{eq:phi_hat_solution} is merely a formal solution of the algebraic equation \eqref{eq:tansformedequation} and is not valid as a proper distributional solution of the problem. In particular, it is ill-behaved at $k = \pm \omega_n$. The correct coordinate-space solution must instead be obtained by applying an adequate integration prescription based on the Cauchy principal value.

The spatial component functions $b_n(x)$ in the coordinate domain can be recovered by performing the inverse Fourier transform:
\begin{equation}
b_n(x) = {\cal F}^{-1}[\hat b_n(k)] = \frac{1}{2\pi} \int_{-\infty}^{\infty} \hat{b}_n(k) e^{ikx} \, dk.
\end{equation}
Substituting the solution for $\hat{b}_n(k)$ into this integral yields:
\begin{equation}
b_n(x) = -\frac{4}{\pi^2 n} \left( \frac{1}{2i} \int_{-\infty}^{\infty} \frac{e^{i k X_{+} }}{k\left(k^2 - \omega_n^2\right)} \, dk - \frac{1}{2i} \int_{-\infty}^{\infty} \frac{e^{i k X_{-} }}{k\left(k^2 - \omega_n^2\right)} \, dk \right),
\end{equation}
where the characteristic boundary distances are defined as $X_{\eta} \equiv x + \eta\frac{L}{2}$ with $\eta = \pm 1$ (abbreviated as $X_{\pm}$).

These integrals feature real simple poles on the integration axis at $k = 0$ and $k = \pm\omega_n$. They can be evaluated analytically in the complex plane by invoking the Cauchy residue theorem using the Cauchy principal value ($PV$) along an appropriately indented contour. Depending on the sign of the exponential arguments $X_{\eta}$, Jordan's lemma dictates closing the integration path in the upper half-plane ($UHP$) for $X_{\eta} > 0$ to ensure exponential decay, or in the lower half-plane ($LHP$) for $X_{\eta} < 0$:
\begin{equation}
\frac{1}{2i} PV\int_{-\infty}^{\infty} \frac{e^{i k X_{\eta} }}{k\left(k^2 - \omega_n^2\right)} \, dk = \frac{\pi}{2}\operatorname{sgn}(X_\eta)\sum_{s}\operatorname{Res}\left(\frac{e^{i k X_{\eta} }}{k\left(k^2 - \omega_n^2\right)},k_s\right) \equiv \operatorname{sgn}(X_\eta) \mathcal{J}_{\eta}(x).
\end{equation}

Consequently, the spatial coordinate space partitions itself into three distinct geometric regions based on the simultaneous activation of these upper and lower half-plane contours. The sign profiles of the arguments $X_{\pm}$ across these spatial domains are summarized in Table~\ref{tab:signs_regions}.

\begin{table}[htbp]
\centering
\caption{Sign profiles of the characteristic parameters $X_{\pm}$ across the three distinct spatial domains.}
\label{tab:signs_regions}
\begin{tabular}{lcc}
\hline
\textbf{Spatial Region} & ${\operatorname{sgn}(X_+)}$ & ${\operatorname{sgn}(X_-)}$ \\ \hline
Left Outer Region: $x < -L/2$           & $-1$ & $-1$ \\
Inner Core Region: $-L/2 \le x \le L/2$ & $+1$ & $-1$ \\
Right Outer Region: $x > L/2$           & $+1$ & $+1$ \\ \hline
\end{tabular}
\end{table}

Computing the sum of the residues at the real poles $k_s \in \{0, \pm\omega_n\}$ under the indented contour prescription yields the explicit regularized parameterization for the spatial source component:
\begin{equation}
\mathcal{J}_{\eta}(x) = \frac{\pi}{2\omega_n^2} \left[ \cos\left(\omega_n X_{\eta}\right) - 1 \right].
\end{equation}
By mapping these components into the spatial Fourier coefficients via the relation
\begin{equation}
b_n(x) = -\frac{4}{\pi^2 n} \left[ \operatorname{sgn}(X_{+})\mathcal{J}_{+}(x) - \operatorname{sgn}(X_{-})\mathcal{J}_{-}(x) \right],
\end{equation}
we can construct the full field profile. Substituting the explicit forms of $\mathcal{J}_{\pm}(x)$ into the odd-mode expansion $\phi(t,x) = \sum_{n=1,3,\ldots} b_n(x) \sin(\omega_n t)$ yields the exact, closed-form spacetime solution:
\begin{equation}\label{eq:phisign}
\phi(t,x) = \sum_{n=1,3,5,\ldots} \frac{-2}{\pi n \omega_n^2} \left[ \operatorname{sgn}(X_{+}) \Big( \cos(\omega_n X_+) - 1 \Big) - \operatorname{sgn}(X_{-}) \Big( \cos(\omega_n X_-) - 1 \Big) \right] \sin(\omega_n t).
\end{equation}
Evaluating coefficients $b_n(x)$ in \eqref{eq:phisign} and applying product-to-sum trigonometric identities yields the final simplified coordinate representation of the spatial profiles:
\begin{equation}
b_n(x) = 
\begin{cases}
-\frac{4}{\pi n\omega_n^2}\sin(\omega_n x)\sin\left(\frac{\omega_n L}{2}\right) & \text{for } x < -\frac{L}{2}, \\[1.5ex]
\frac{4}{\pi n\omega_n^2} \left[1-\cos(\omega_n x)\cos\left(\frac{\omega_n L}{2}\right) \right] & \text{for } -\frac{L}{2} \le x \le \frac{L}{2}, \\[1.5ex]
\frac{4}{\pi n\omega_n^2}\sin(\omega_n x)\sin\left(\frac{\omega_n L}{2}\right) & \text{for } x > \frac{L}{2}.
\end{cases}
\label{eq:phi_n_piecewise2}
\end{equation}

The requirement of spatial compact support dictates that the solution must vanish identically in the exterior domain where $|x| > \frac{L}{2}$. According to the exterior definitions in \eqref{eq:phi_n_piecewise2}, this condition is satisfied if and only if the argument $\omega_n \frac{L}{2}$ constitutes an integer multiple of $\pi$. Recalling that $\omega_n = \frac{2\pi n}{T}$, the parameter constraint can be rewritten as $\omega_n\frac{L}{2} = n \pi \frac{L}{T}$. This establishes that the physical parameters of the system must satisfy the following   condition:
\begin{equation}
L = NT, \quad \text{where } N \in \mathbb{N}^+.
\label{eq:quantization}
\end{equation}

Physically, equation \eqref{eq:quantization} implies that the spatial width of the compact oscillon must be a strict integer multiple of its temporal period. The fundamental case $N=1$ describes an isolated, individual compact oscillon. Higher-order configurations ($N > 1$) represent a regular, synchronized train of compact oscillons residing on contiguous, non-overlapping spatial intervals whose adjacent boundaries touch. Applying this condition yields $\cos\left(\frac{\omega_n L}{2}\right) = (-1)^{nN}$.
By expressing the spatial profiles as $b_n(x) = B^{(N)}_n(x) \Pi(x)$, the full analytical Fourier representation of the compact oscillon across all space is given by
\begin{equation}\label{Fourier_oscillon}
\phi(t,x) = \sum_{n=1,3,5,\ldots}^{\infty} B^{(N)}_n(x) \Pi(x) \sin\left(\omega_n t\right),
\end{equation}
where $\Pi(x)$ is the rectangular window function defined in \eqref{window} that explicitly enforces the spatial boundaries of the compacton, and
\begin{equation}\label{eq:Bncoef}
B^{(N)}_n(x) \equiv \frac{4}{\pi n\omega_n^2} \Big[1 - (-1)^{nN} \cos\left(\omega_n x\right) \Big].
\end{equation}

\begin{figure}[h!]
\includegraphics[width=0.45\linewidth]{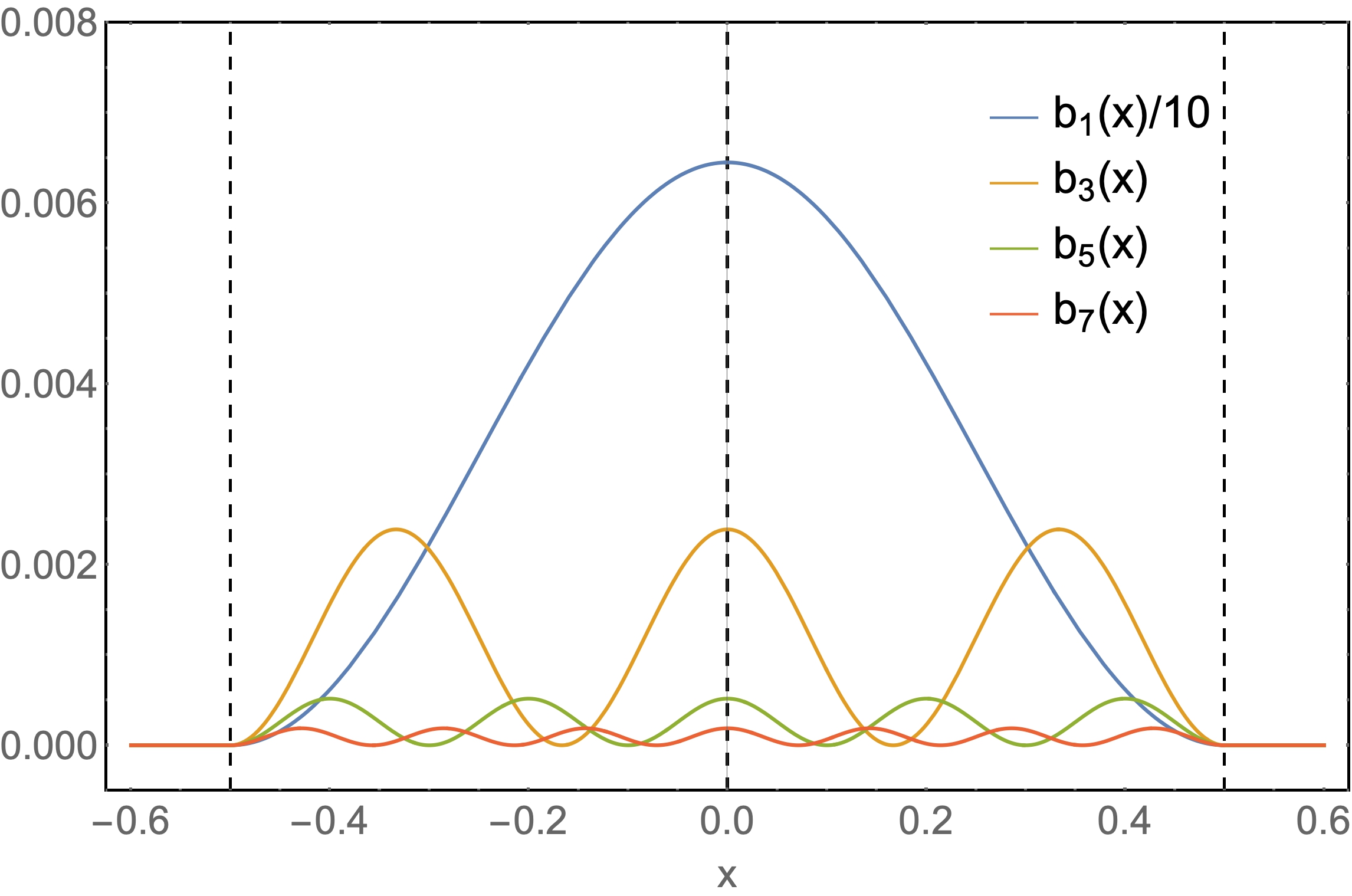}
\caption{Spatial profiles of the coefficients $b_n(x)$ for $N=1$ and odd modes $n=1,3,5,7$. For enhanced visibility, $b_1(x)$ is rescaled by a factor of $10$.}\label{fig:bn0}
\end{figure}

The coefficients $b_n(x) = B^{(1)}_n(x) \Pi(x)$ are shown in Fig.~\ref{fig:bn0}. The maximum of $b_1(x)$ occurs at $x = 0$ with a value of $b_1(0)=\frac{2T^2}{\pi^3}$. More generally, the $n$-th order coefficient $b_n(x)$ exhibits $n$ peaks, whose amplitudes scale as $n^{-3}$.

\begin{figure}[h!]
\subfigure[]{\includegraphics[width=0.31\linewidth]{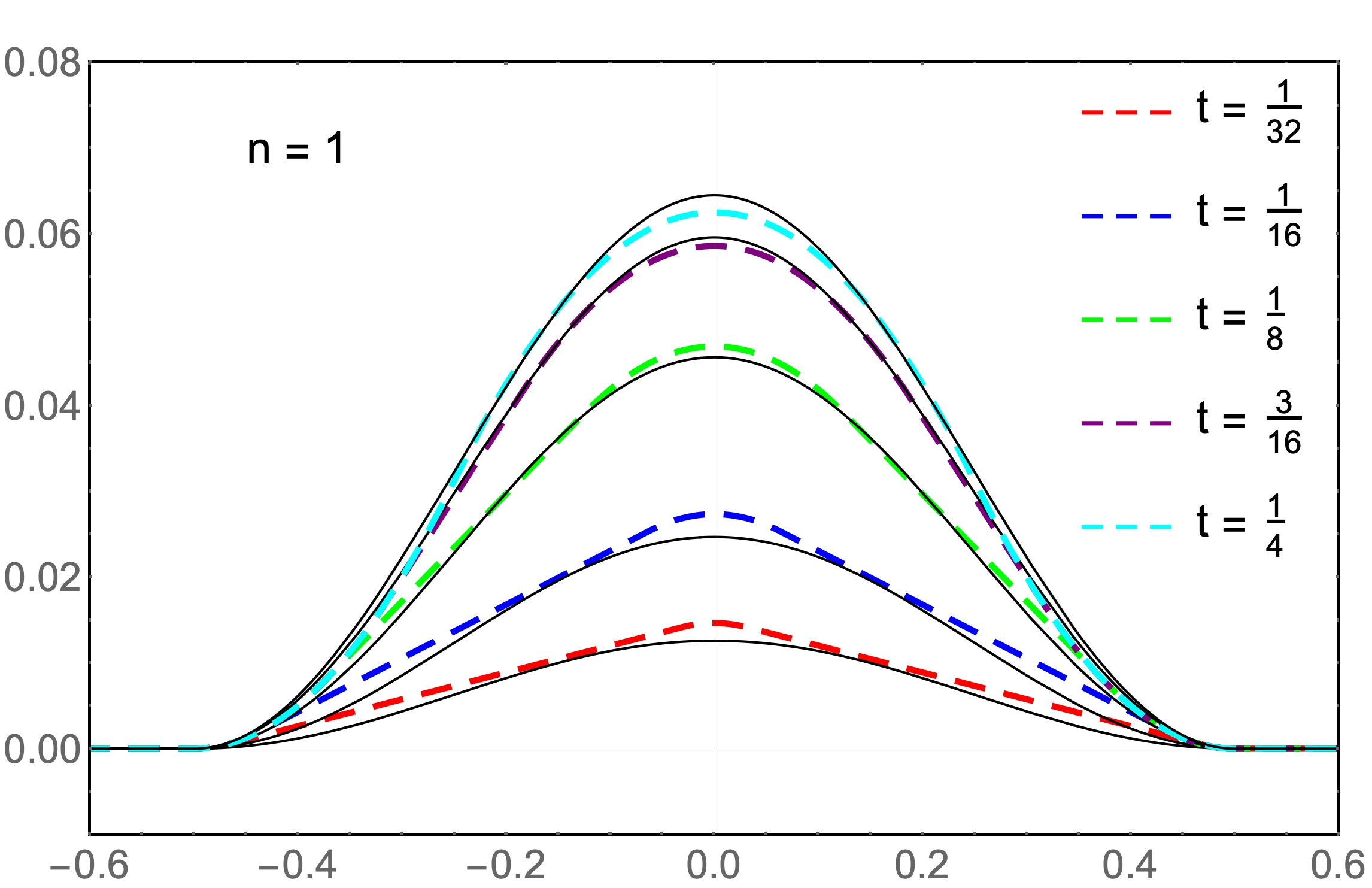}\label{fig:p1}}
\subfigure[]{\includegraphics[width=0.31\linewidth]{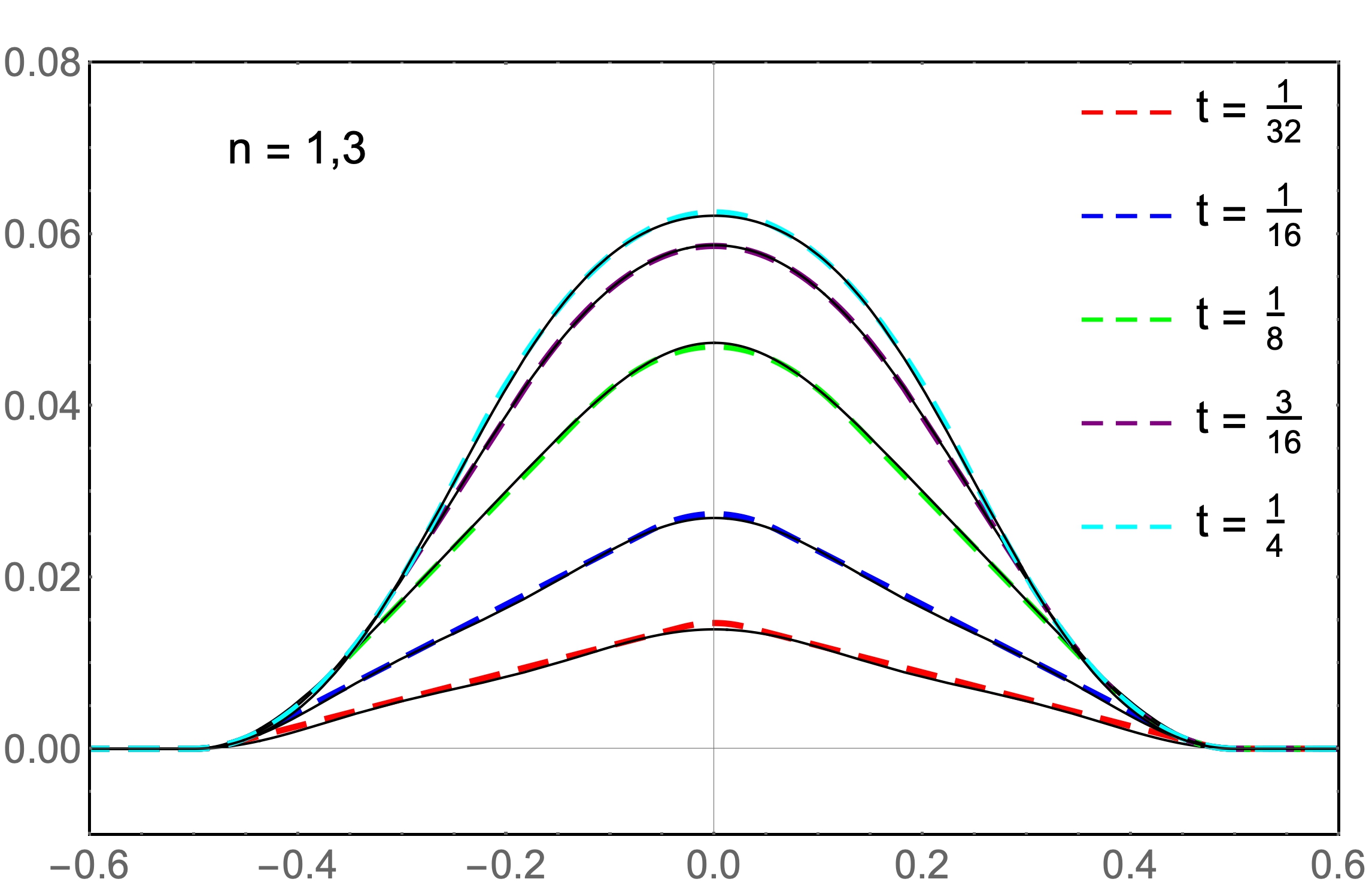}\label{fig:p3}}
\subfigure[]{\includegraphics[width=0.31\linewidth]{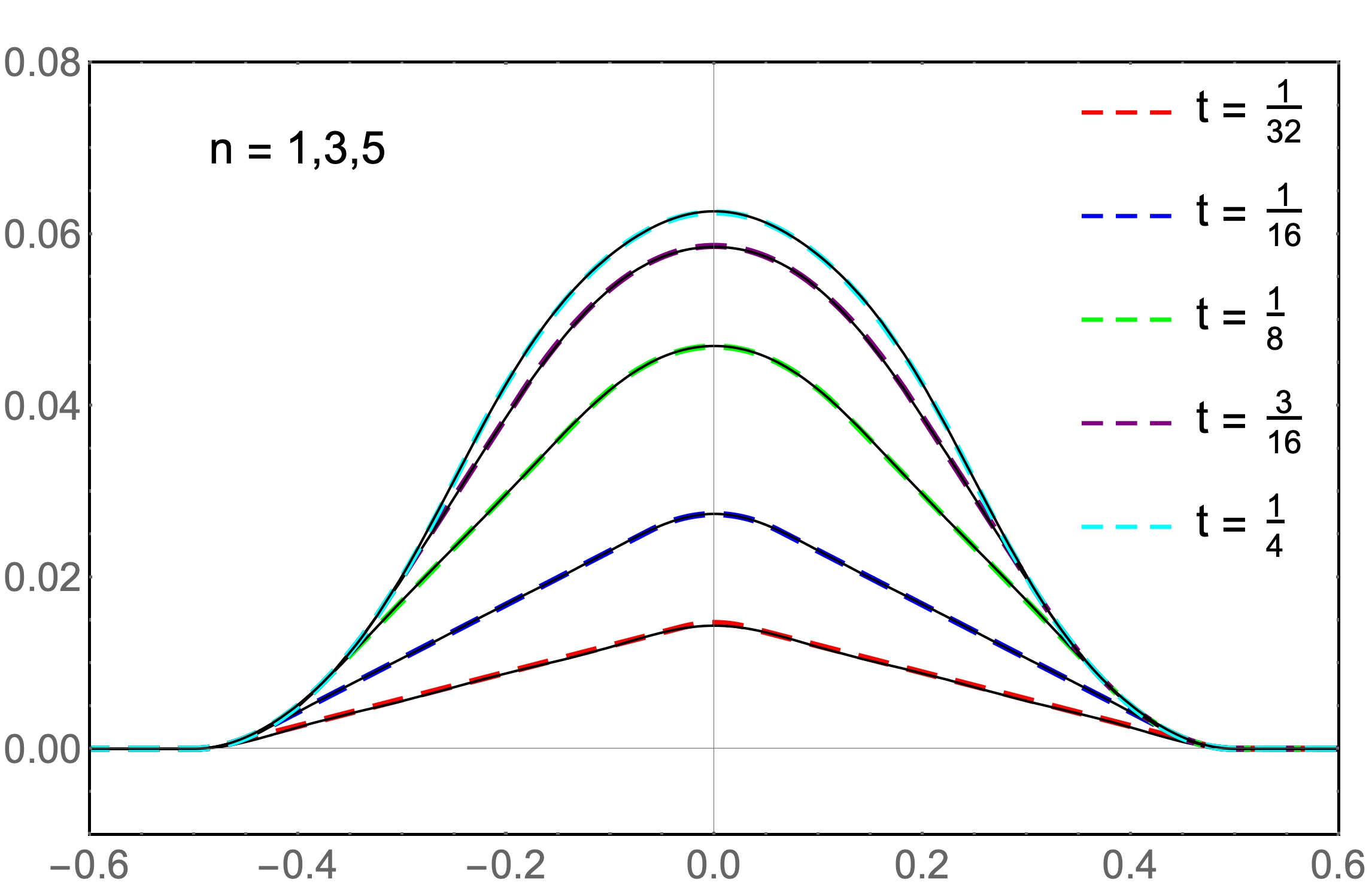}\label{fig:p5}}
\caption{Snapshots of the exact oscillon (dashed line) and its Fourier approximations containing terms with (a) $n=1$, (b) $n=1,3$, and (c) $n=1,3,5$.}\label{exactvsfourier}
\end{figure}
\begin{figure}[h!]
{\includegraphics[width=0.9\linewidth]{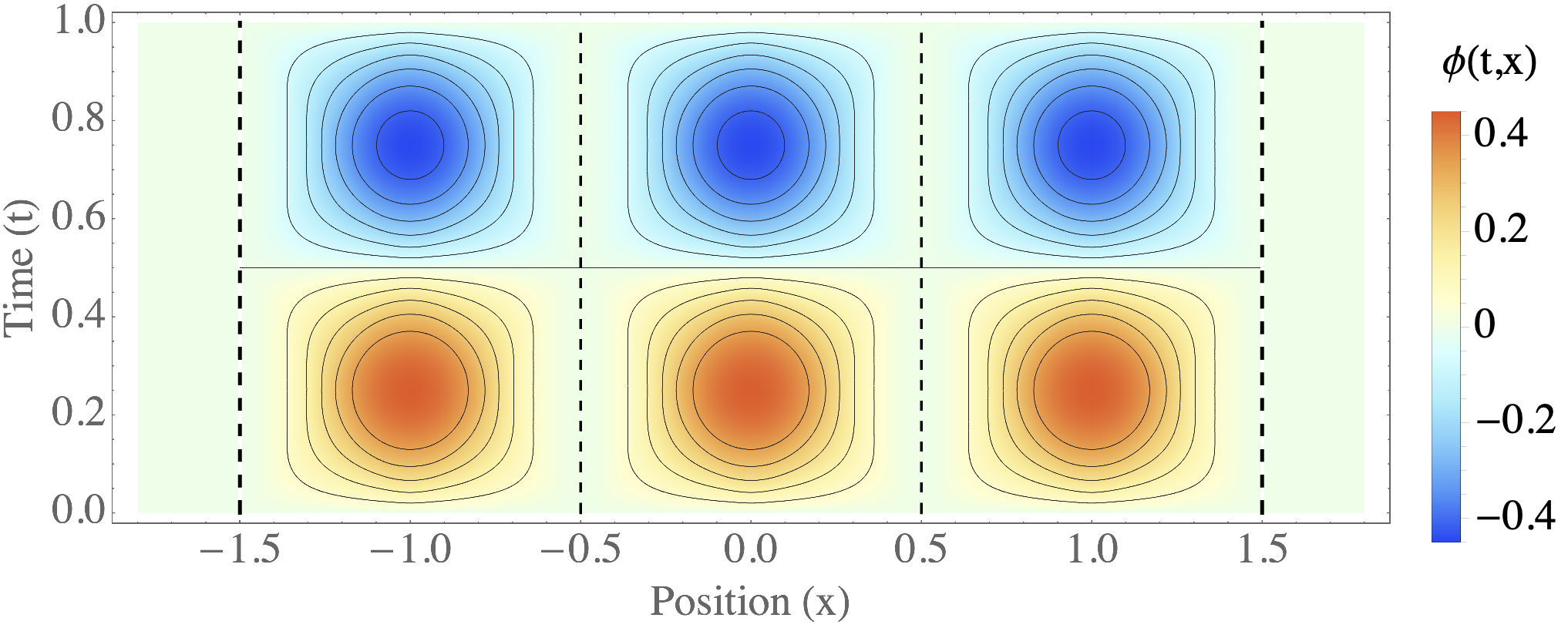}\label{fig:multi}}
\caption{Multioscillon $N=3$.}\label{multioscillon}
\end{figure}

In Fig.~\ref{exactvsfourier}, we present several snapshots comparing the temporal evolution of the exact oscillon solution for $N=1$ against its spectral reconstructions. Specifically, Figs.~\ref{fig:p1}, \ref{fig:p3}, and \ref{fig:p5} display the profiles obtained by retaining one, two, and three terms in the Fourier series expansion, respectively. The rate of numerical convergence for this series is remarkably rapid; a truncation containing merely the first three harmonic components ($n=1, 3, 5$) already provides an excellent approximation of the exact, fully non-linear oscillon configuration. In Fig.~\ref{multioscillon}, we present a representative spacetime profile of a multi-oscillon configuration composed of $N=3$ individual segments, showcasing the localization of the field boundaries.

The principal advantage of this spectral approach lies in the drastic reduction of the number of localized, piecewise boundary matchings required to establish the profile. In conventional spatial matching techniques, constructing the exact solution necessitates splicing together a multitude of local configurations across internal light-cone characteristics. Such procedures introduce significant analytical overhead due to the algebraic complexity of enforcing matching and continuity conditions across these multiple surfaces. In contrast, the Fourier framework elegantly reformulates the problem by partitioning the global domain into only two fundamental regimes: a non-trivial internal core solution and a homogeneous external vacuum state.

\subsection{Fourier analysis of the initial configuration}

For an isolated single oscillon with $N = 1$ ($L = T$), the internal spatial profiles are given by 
\[
b_n(x) = B_n(x)\Pi(x) = \frac{4}{\pi n \omega_n^2} \left[1 + \cos(\omega_n x)\right]\Pi(x).
\] 
Here, the upper index in $B_n^{(1)}(x)$ has been omitted for brevity [see \eqref{eq:Bncoef}]. Consequently, the spacetime evolution of the field $\phi(t,x)$ is expressed as:
\[
\phi(t,x)=\sum_{n=1,3,\ldots}^{\infty}B_n(x)\Pi(x)\sin(\omega_n t).
\]
where $\Pi(x)$ is a window function \eqref{window}.
The initial velocity distribution of the field, $\psi(x) \equiv \partial_t\phi(t,x)\big|_{t=0}$, can be expressed as the infinite series:
\begin{align}
\psi(x) = \sum_{n=1,3,5,\ldots}^{\infty}\omega_n B_n(x) \Pi(x)=\frac{2T}{\pi^2} \sum_{n=1,3,5,\ldots}^{\infty} \frac{1}{n^2} \left[1 + \cos\left(\omega_n x\right) \right] \Pi(x). \label{eq:psi_series_final_0}
\end{align}
Within the core spatial domain $|x| \le \frac{T}{2}$, this expansion can be evaluated analytically by invoking the standard Fourier series identities for odd integers 
\[
\sum_{n=1,3,5,\ldots}^{\infty} \frac{1}{n^2} = \frac{\pi^2}{8}\qquad\text{and}\qquad \sum_{n=1,3,5,\ldots}^{\infty} \frac{\cos(n\theta)}{n^2} = \frac{\pi^2}{8} - \frac{\pi}{4}|\theta| \qquad \text{for}\qquad |\theta| \le \pi.
\] 
By introducing the angular parametrization $\theta = \frac{2\pi x}{T} = \omega_1 x$, the infinite summation reads:
\begin{equation}\label{eq_triangle}
\psi(x) = \left( \frac{T}{2} - |x| \right) \Pi(x),
\end{equation}
which explicitly recovers the symmetric triangular profile characteristic of the initial compacton velocity field.

To extract the physical consequences of this formulation regarding the steady character of the compact borders, we take the continuous spatial Fourier transform of the triangular distribution:
\begin{equation}
\hat{\psi}(k) = \int_{-\infty}^{\infty} \psi(x) e^{-ikx} \, dx = \int_{-T/2}^{T/2} \left( \frac{T}{2} - |x| \right) e^{-ikx} \, dx.
\end{equation}
Evaluating this symmetric triangle integral yields the continuous wavenumber spectrum:
\begin{equation}\label{eq:psi_k_final}
\hat{\psi}(k) = \frac{4 \sin^2\left(\frac{kT}{4}\right)}{k^2}.
\end{equation}

The asymptotic power law $\mathcal{O}(k^{-2})$ is the characteristic signature of a continuous but non-differentiable ($C^0$-class) function, such as the triangular distribution defined in Eq.~\eqref{eq_triangle}. This behavior stems from the derivative property of the Fourier transform, where an $m$-th derivative containing sharp discontinuities yields a spectral decay of $|k|^{-(m+1)}$ as $|k| \to \infty$. Notably, the absence of a slower $\mathcal{O}(k^{-1})$ regime in Eq.~\eqref{eq:psi_k_final} rules out the formation of localized $\delta$-shocks or singular pressure fronts at the interface. This decay rate ensures that the boundary profiles transition continuously into the unperturbed vacuum without introducing unphysical derivative discontinuities into the field equations.

Projecting this global velocity profile onto the discrete higher-order internal odd harmonics of the oscillon yields:
\begin{equation}\label{psiondiscretespectum}
\hat{\psi}(\omega_n)= \frac{4}{\omega_n^2} \sin^2\left(\frac{n\pi}{2}\right)=\frac{4}{\omega_n^2}.
\end{equation}
This result is consistent with the spectral expansion of the velocity field obtained from \eqref{eq:psi_series_final_0}. By applying the spatial Fourier transform to the mode expansion and invoking the convolution theorem, we obtain:
\begin{equation}\label{eq:convolution}
\mathcal{F}\big[ B_n(x)\Pi(x)\big](k) = \frac{1}{2\pi}\int_{-\infty}^{\infty}\hat B_n(k')\hat\Pi(k-k')\,dk',
\end{equation}
where the coefficients
\begin{equation}\label{eq:spatialtransform}
\hat B_n(k) = \frac{8}{n\omega_n^2}\Big[\delta(k) + \frac{1}{2}\delta(k-\omega_n) + \frac{1}{2}\delta(k+\omega_n)\Big]
\end{equation}
represent the Fourier transforms of $B_n(x)$, and the function $\hat \Pi(k)$ is given in \eqref{eq:transformPi}.
Evaluating this convolution explicitly yields the continuous Fourier transform of the global velocity field as a sum over the discrete spectrum:
\begin{equation}\label{eq:psi_k_final_2}
\hat{\psi}(k) =\sum_{n=1,3,\ldots}^{\infty}\frac{4}{\pi n\omega_n}\left[2\frac{\sin\left(\frac{kT}{2}\right)}{k}+\frac{\sin\left(\frac{(k-\omega_n)T}{2}\right)}{k-\omega_n}+\frac{\sin\left(\frac{(k+\omega_n)T}{2}\right)}{k+\omega_n}\right].
\end{equation}
When sampling Eq.~\eqref{eq:psi_k_final_2} at the discrete frequencies $k = \omega_m$, the first and third sinc terms vanish identically because their arguments evaluate to odd-integer multiples of $\pi$, where the sine function has roots. For the same reason, the middle sinc term vanishes for all off-diagonal components where $n \neq m$. The only non-vanishing contribution arises from the resonant denominator where $\omega_m - \omega_n = 0$, which evaluates to $\frac{T}{2}\delta_{mn}$ in the limit. This immediately collapses the infinite summation to the single diagonal term:
\begin{equation}
\hat{\psi}(\omega_m) = \frac{4}{\pi m \omega_m} \left( \frac{T}{2} \right) = \frac{4}{\omega_m^2}.
\end{equation}

This exact agreement between the sampled continuous spectrum and the discrete modal coefficients carries profound physical consequences. It demonstrates that the infinite summation of truncated standing wave modes converges  to the global geometric profile without inducing any spectral mismatch or numerical ringing (Gibbs phenomenon) at the boundaries. Because the amplitudes of the higher-order odd harmonics decay at the exact rate ($\mathcal{O}(\omega_m^{-2})$) required to reconstruct the sharp slope of the initial triangular distribution, the modes remain perfectly phase-locked. This precise spectral alignment ensures that the global, localized compact object is entirely self-contained; the internal discrete harmonic channels support the sharp boundary profile perfectly, preventing any energy leakage into the surrounding vacuum and reinforcing the character of the system as a self-sustaining dynamical trap.

\subsection{Oscillon's spectrum}

The complete spectrum of a single $N=1$ compact oscillon is given by the double Fourier transform:
\begin{align}
\hat\phi(\omega',k)&=\int_{-\infty}^{\infty} dt \int_{-\infty}^{\infty} dx \,\phi(t,x) e^{i(\omega' t - kx)} \\
&=\sum_{n=1,3,\ldots}^{\infty} \left( \int_{-\infty}^{\infty} \sin(\omega_n t) e^{i\omega' t} \, dt \right) \left( \int_{-\infty}^{\infty} B_n(x)\Pi(x) e^{-ikx} \, dx \right),
\end{align}
where $B_n(x)=\frac{4}{\pi n\omega_n^2}(1+\cos(\omega_n x))$. The temporal Fourier transform yields a pair of Dirac delta functions localizing the energy at the driving frequencies:
\begin{equation}
{\cal F}[\sin(\omega_n t)] = i\pi \Big[ \delta(\omega' + \omega_n) - \delta(\omega' - \omega_n) \Big].
\end{equation}
Concurrently, the spatial Fourier transform can be evaluated via the convolution theorem or direct integration over the compact domain $[-T/2, T/2]$ (since $L=T$ for $N=1$), which results in a linear combination of frequency-shifted sinc functions. Combining these components yields the full analytical expression for the spacetime spectrum:
\begin{align}
\hat\phi(\omega',k)&=\sum_{n=1,3,\ldots}^{\infty}\frac{4}{in\omega_n^2}\left(2\frac{\sin\left(\frac{kT}{2}\right)}{k}+\frac{\sin\left(\frac{k-\omega_n}{2}T\right)}{k-\omega_n}+\frac{\sin\left(\frac{k+\omega_n}{2}T\right)}{k+\omega_n}\right)
\Big[\delta(\omega'-\omega_n)-\delta(\omega'+\omega_n)\Big].\label{eq:completespectrum}
\end{align}
Notably, the temporal spectrum is entirely discrete, consisting solely of the fundamental frequency and its odd higher harmonics. In contrast, the spatial spectrum is continuous for each harmonic, characterized by a smooth, singularity-free profile that is uniquely determined by the spatial distribution of the localized bound state.

\subsection{First-mode ($n=1$) approximation}

In Section~\ref{sec:densities}, we analyzed the energy and linear momentum density distributions for the exact oscillon configurations. By examining the behavior of these densities in the left ($x < 0$) and right ($x > 0$) domains of the oscillon, we observed that their respective centers of energy (or mass) follow periodic trajectories, alternating between approach and separation. This behavior raises the question of whether the oscillon can be interpreted as a bound state of two distinct constituents. In this section, we investigate this possibility by analyzing the Fourier modes of the exact oscillon solutions.

First, we observe from the complete spectrum of the exact compact oscillon \eqref{eq:completespectrum} that it comprises a fundamental frequency $\omega = \frac{2\pi}{T}$ alongside an infinite tower of odd harmonics $\omega_n$ for $n = 3, 5, 7, \ldots$. The absence of non-harmonic frequencies provides strong evidence that the oscillon is a fundamental, single entity rather than a composite system. To explore this further, we isolate the first fundamental mode ($n = 1$) and analyze the corresponding dynamics of its energy and momentum densities, as well as the behavior of its centers of mass.

The first-mode ($n=1$) approximation of a single $N=1$ compact oscillon is expressed as
\begin{equation}
\phi_1(t,x)=\frac{4}{\pi\omega^2}\Big[1+\cos(\omega x)\Big]\sin(\omega t),\qquad |x|\le \frac{T}{2}
\label{eq:phiaprox}
\end{equation}
where $\omega \equiv \omega_1 = \frac{2\pi}{T}$. The snapshots representing the profile of this function are shown in Fig.~\ref{fig:p1}. Within the domain $|x| \le \frac{T}{2}$, the energy and momentum densities associated with this approximation take the form:
\begin{align}
\mathcal{H}_1 &= \frac{16}{\pi^2\omega^2} \cos^2\left(\frac{\omega x}{2}\right) \left[1+\cos(\omega x)\cos(2\omega t)+\frac{\pi}{2} \big| \sin(\omega t)\big|\right], \label{eq:Haproc} \\
\mathcal{P}_1 &= \frac{8}{\pi^2\omega^2} \left[\sin(\omega x) + \frac{1}{2}\sin(2\omega x) \right]\sin(2\omega t).\label{eq:Paprox}
\end{align}

Note that $\mathcal{P}_1$ is an odd function of $x$, and thus the total momentum of the oscillon approximated by \eqref{eq:phiaprox} vanishes upon integration over a symmetric spatial domain. On the other hand, the total energy associated with the first mode, $E_1(t) = \int_{-T/2}^{T/2} \mathcal{H}_1 \, dx$, is not conserved; it is an explicit function of time given by:
\begin{equation}
E_1(t) = \frac{T^3}{\pi^4}\Big[2+\cos(2\omega t)+\pi|\sin(\omega t)|\Big].
\end{equation}

\begin{figure}[h!]
\subfigure[]{\includegraphics[width=0.45\linewidth]{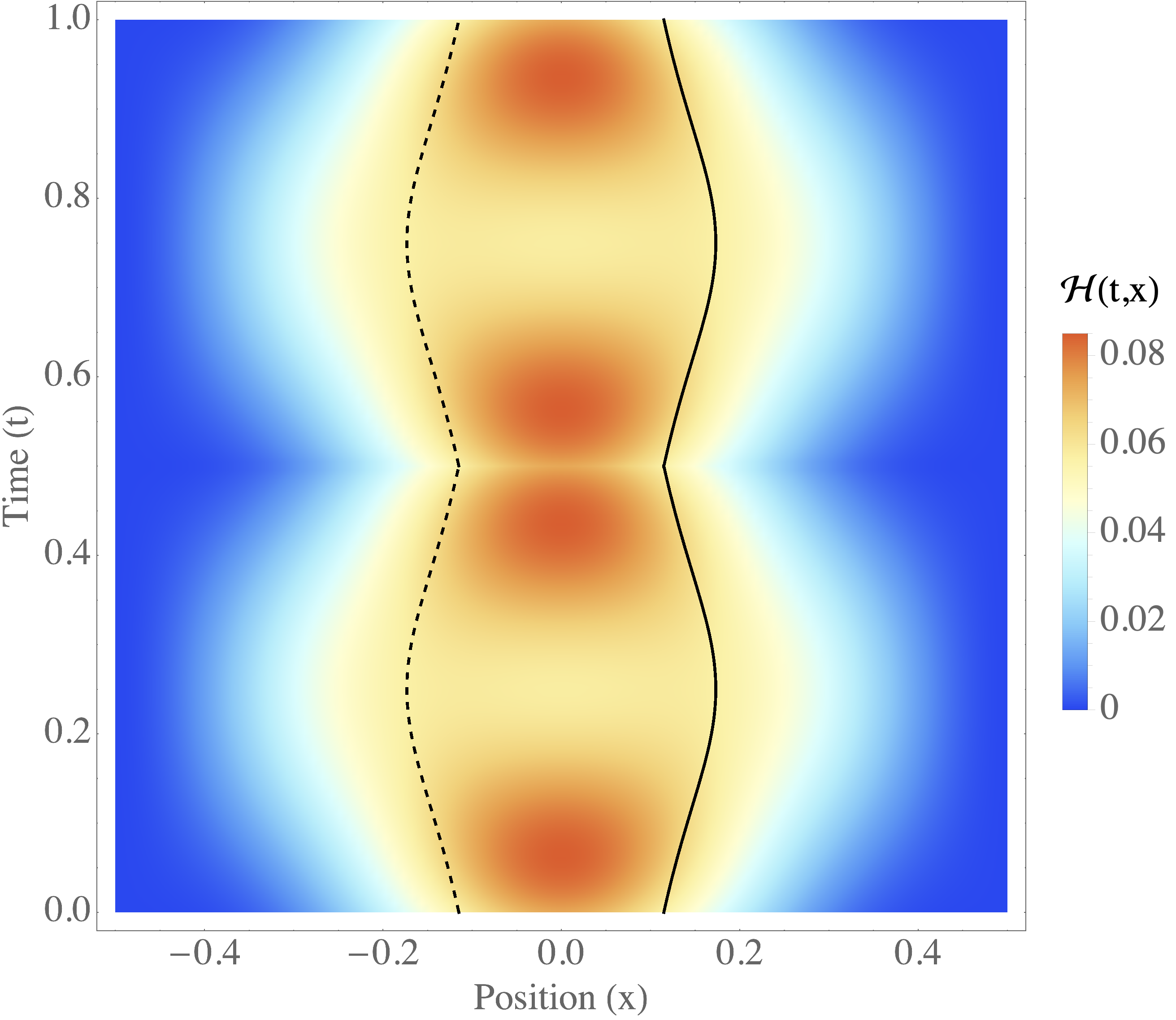}\label{fig:traj-energy-firstmode}}\;\;
\subfigure[]{\includegraphics[width=0.45\linewidth]{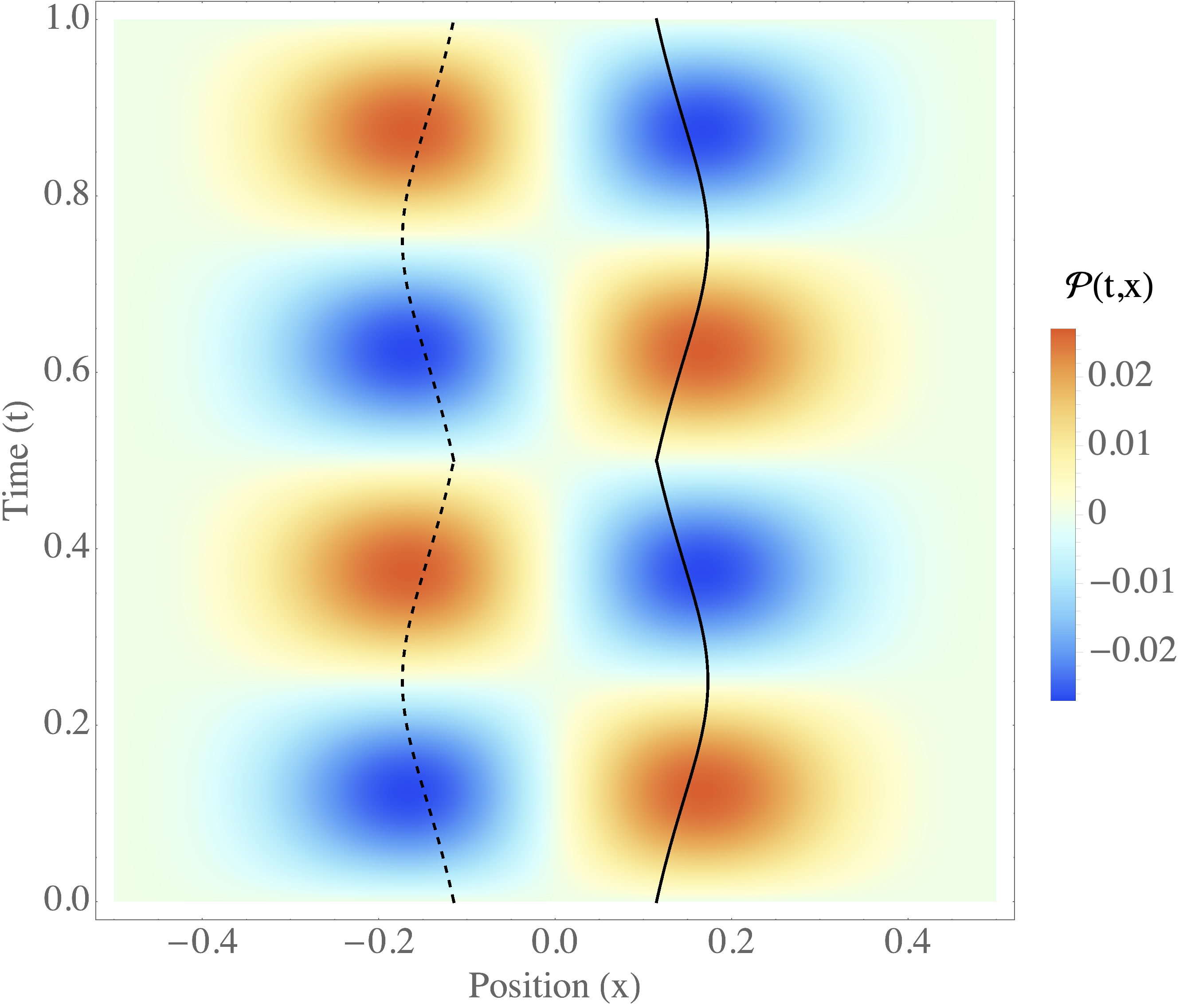}\label{fig:traj-momentum-firstmode}}
\caption{The first mode $n=1$ approximation for the exact oscillon. Spatiotemporal trajectories of the centers of mass for the left (dashed line) and right (solid line) subsystem components overlaid on the background profiles of (a) the energy density $\mathcal{H}_1(t,x)$ and (b) the linear momentum density $\mathcal{P}_1(t,x)$.}\label{fig:trajectories-densities}
\end{figure}

The expression $E_1(t)$ is a continuous but non-smooth function, exhibiting sharp spikes at integer multiplicities of the oscillon half-period, $t = jT/2$ for $j \in \mathbb{Z}$. This lack of energy conservation is an artifact of the chosen frequency cutoff. The energy $E_1(t)$ is strictly associated with the single localized oscillon mode, deviating significantly from the true conserved total energy $E_{\text{tot}} = \frac{T^3}{24}$ in the vicinity of these half-period configurations.
Following definitions \eqref{eq:cm_definitions} we introduce mass center associated with first mode contribution taking $E_L(t)=E_R(t)=\frac{1}{2}E_1(t)$. This results in
\begin{equation}
X^{(1)}_R(t)=-X^{(1)}_L(t)=\frac{T^2}{4\pi^2}\frac{(\pi^2-8)\cos(2\omega t)+(\pi^2-4)(2+\pi|\sin(\omega t)|)}{2+\cos(2\omega t)+\pi|\sin(\omega t)|}.\label{eq:trajectoriesfirstmode}
\end{equation}
The trajectories \eqref{eq:trajectoriesfirstmode} are plotted  against the background of the energy density (Fig.~\ref{fig:traj-energy-firstmode}) and the momentum density (Fig.~\ref{fig:traj-momentum-firstmode}). These figures show that the qualitative behavior of the trajectories, as well as the spatial distributions of the energy and momentum densities discussed in Section~\ref{sec:densities} (Fig.~\ref{fig:traj-energy} and \ref{fig:traj-momentum} ), are well reproduced already by the first mode alone. This agreement allows us to conclude that a compact, exact oscillon in the signum-Gordon model behaves as a single coherent structure rather than a bound system consisting of two separate components.

\subsection{Comparison with oscillons in models with analytic potentials}

There is extensive literature on oscillons and breathers (or quasi-breathers) in classical field-theoretic models with analytic (smooth) potentials. To highlight the key characteristics of signum-Gordon oscillons and compare them with those in analytic potentials, we consider the seminal paper \cite{PhysRevD.78.025003}, which summarizes the main features of such configurations.

Small-amplitude oscillons were investigated by Fodor et al. \cite{PhysRevD.78.025003} in the context of a nonlinear wave equation with an analytic potential whose field derivative expands as $V'(\phi)=\phi+\sum_{k=2}^{\infty}g_k\phi^k$ (denoted as $U'$ in their paper). In this parametrization, the perturbative mass is normalized to unity, $m^2=V''(0)=1$. In contrast, the signum-Gordon potential $V(\phi)=\vert\phi\vert$ lacks well-defined derivatives at $\phi=0$. While a series expansion can formally be introduced using generalized functions---where $V'$ is the signum function and $V''$ is proportional to the Dirac delta distribution---this approach fails to yield a physically meaningful perturbative mass, formally suggesting an infinite value instead. A more promising alternative relies on analyzing the dispersion relation of the signum-Gordon field \cite{Streibel:2026yxv}.

The expansion parameter $\epsilon$ controls the oscillon amplitude (where $\phi \to 0$ as $\epsilon \to 0$) and sets the characteristic timescale governing the oscillation frequency, $\omega^2(\epsilon) = 1 + \sum_{k=1}^{\infty} \epsilon^k \omega_k$. The value $\omega = 1$ corresponds to the small-amplitude threshold determined by the mass of the scalar field. As $\epsilon \to 0$, the spatial extent of these smooth configurations increases; thus, smaller amplitudes correspond to broader, less localized oscillons. Comparing these universal properties with the features of signum-Gordon oscillons reveals several stark differences. Setting the signum-Gordon oscillon period to $T = \epsilon$, we obtain its characteristic amplitude $A = \frac{\epsilon^2}{16}$, size $L = \epsilon$, and frequency $\omega(\epsilon) = \frac{2\pi}{\epsilon}$. This demonstrates that small-amplitude signum-Gordon oscillons remain spatially compact, in sharp contrast to their counterparts in analytic potentials. Furthermore, there is no frequency threshold; the frequency diverges to infinity as $\epsilon \to 0$.

Another crucial issue is the existence of genuine, stable oscillons. In the perturbative approach, the scalar field is represented as a formal series, $\phi = \sum_{k=1}^{\infty} \epsilon^k \phi_k$. Solving the nonlinear wave equation perturbatively requires solving an infinite tower of coupled differential equations order by order. At the lowest orders, a universal master equation holds for analytic potentials \cite{PhysRevD.78.025003}. Solutions to this equation are referred to as quasi-breathers \cite{PhysRevD.74.124003}. A major limitation of quasi-breather solutions is their spatial localization: they consist of a core containing most of the energy density, accompanied by non-vanishing spatial tails. While the core accurately models the central dynamics of oscillons, the total energy stored in these infinite spatial tails is unbounded, rendering the complete mathematical solution unphysical. Furthermore, as shown in \cite{PhysRevLett.58.747}, true small-amplitude periodic states do not exist in $\phi^4$ field theory, implying that the perturbative expansion does not converge and must be treated as an asymptotic series. The analysis in \cite{PhysRevD.74.124003} demonstrates that oscillons---long-lived, metastable states closely approximated by quasi-breathers---are never strictly localized or stable, but perpetually radiate energy at a small rate.

The exact signum-Gordon compact oscillons evade the theorem regarding the non-existence of spatially localized, small-amplitude periodic states. Moreover, as demonstrated in Eq.~\eqref{Fourier_oscillon}, the representation of the exact oscillon as a Fourier series is perfectly sound; the entire series converges uniformly to the oscillon profile. This example demonstrates, therefore, that the non-analyticity of the signum-Gordon potential allows strictly compact, exactly periodic solutions to evade the standard small-amplitude non-existence results.

A qualitatively similar behavior has also been observed numerically in $CP^N$ models with V-shaped potentials \cite{Klimas:2017eft,Colaco:2026aua}, in which compact solutions have been constructed for odd $N$. Such models are formulated in terms of $N$ complex scalar fields. For odd $N\equiv 2\,\ell+1$, compactons can be constructed using an ansatz of the form $u_m \propto f(r)\,Y_{\ell m}\,e^{i\,\omega\, t}$, with $m=-\ell,\ldots,\ell$, where $Y_{\ell m}$ are spherical harmonics, $\omega$ is the angular frequency, and $f(r)$ is a common radial profile function. In the high-frequency regime, both the maximum value of $f$ and the compacton radius become very small and tend to decrease further as $\omega$ increases. This suggests that the simultaneous small-field, high-frequency, and shrinking-support behavior may be a more general feature of models with V-shaped potentials.

Finally, as discussed in the previous section, there is no evidence to suggest that our oscillon is a bound state of other oscillons. First, its power spectrum contains exclusively harmonic frequencies. Second, following the mechanism described in \cite{PhysRevLett.134.081601}, a bound state of two unmodulated oscillons produces an oscillon exhibiting amplitude modulation, whereas the fundamental signum-Gordon oscillon shows no amplitude modulation whatsoever. Nevertheless, we cannot rule out this possibility for compact perturbed signum-Gordon oscillons, which occasionally decay by emitting smaller compact oscillons. Addressing this open question will require further, more detailed investigation.

\section{Swaying compact oscillon}
\label{sec:Fourierswaying}

\subsection{Basic characteristics}

In this section, we generalize the Fourier decomposition of a compact oscillon to the case of a periodically moving boundary.

Swaying oscillons constitute a class of periodic solutions that exhibit periodic motion of their compact support in addition to the periodic variation of the field's magnitude. The simplest oscillon of this type, featuring boundaries that move uniformly along zigzag worldlines, was introduced in \cite{Arodz:2011zm}. Later, these configurations were generalized to cases with non-uniform boundary motion \cite{Swierczynski:2016ivq}. The left and right boundaries of the swaying oscillon are characterized by periodic zigzag worldlines defined as $x_L(t) = -\frac{T}{2} + y(t)$ and $x_R(t) = \frac{T}{2} + y(t)$, where $T$ represents the mean width of the support and $y(t)$ describes the collective periodic sway. Here, we explicitly set $L = T$ at the outset for this single-oscillon configuration to simplify the resulting expressions.

The simplest oscillon of this kind possesses a displacement trajectory $y(t) = v_0 \tau(t)$, where the saw-tooth function $\tau(t)$ is represented by the infinite series:
\begin{equation}\label{tau}
\tau(t) = \sum_{m=1,3,5,\ldots}^{\infty}\frac{2T}{(\pi m)^2}\left(1-\cos(\omega_m t)\right), \quad \text{with} \quad \omega_m = \frac{2\pi m}{T}.
\end{equation}
Its time derivative corresponds to $\frac{d\tau}{dt} = \Sigma(t)$, where $\Sigma(t) = \text{sgn}(\sin(\omega t))$ is the square-wave function defined previously in \eqref{sigma}. Consequently, within a single full period $t \in [0, T]$, the boundary displacement function $y(t)$ takes the explicit piecewise form:
\begin{equation}\label{eq:y-trajectory}
y(t) = \begin{cases} 
v_0 t & \text{for } 0 \le t \le \frac{T}{2}, \\[10pt]
v_0 (T - t) & \text{for } \frac{T}{2} < t \le T.
\end{cases}
\end{equation}
The asymmetry parameter $v_0 \neq 0$ is introduced directly into the initial velocity profile:
\begin{equation}
\psi(x) = 
\begin{cases}
\dfrac{1}{1+v_0} \left(x + \dfrac{T}{2}\right) & \text{for } -\dfrac{T}{2} \le x \le v_0 \dfrac{T}{2}, \\[12pt] 
\dfrac{1}{1-v_0} \left(\dfrac{T}{2} - x\right) & \text{for } v_0 \dfrac{T}{2} < x \le \dfrac{T}{2}. 
\end{cases}
\end{equation}
As a consequence of this spatial asymmetry, the system exhibits a global \emph{swaying motion}, with the compact field boundaries oscillating dynamically in time. Consequently, the half-period sign-reversal symmetry condition \eqref{cond3} no longer holds. Crucially, this swaying motion is not a slow, adiabatic modulation, but rather a \textit{phase-locked and fully synchronized} mode driven by the fundamental fast oscillations of the underlying field.

\subsection{Derivation of spatial profile functions}
Because the swaying motion shares the exact periodicity of the field, we can expand $\phi(t,x)$ using a strict Fourier series with time-independent spatial profiles $a_n(x)$ and $b_n(x)$ over the compact domain $[x_L(t), x_R(t)]$. The mode expansion reads:
\begin{equation}
\phi(t,x) = \sum_{n=1}^{\infty} \Big[ a_n(x)\cos(\omega_n t) + b_n(x)\sin(\omega_n t) \Big], \qquad x_L(t)\le x\le x_R(t).
\end{equation}
Note that here the summation index $n$ runs over \textbf{all positive integers} (both odd and even). The inclusion of the even harmonics is a direct mathematical consequence of the broken half-wave symmetry \eqref{cond3} induced by the asymmetric, moving boundaries.

The signum driving force represents a periodic switching window. In the co-moving frame where the core domain of width $T$ sways according to the trajectory $y(t)$, the window function can be defined explicitly using the standard Heaviside step function $\theta(\xi)$:
\begin{equation}\label{eq:pi_heaviside}
\Pi(t,x) = \theta\left(x +\frac{T}{2} - y(t)\right) - \theta\left(x -\frac{T}{2} -y(t)\right).
\end{equation}

In this spatial configuration, the signum driving force is parameterized via the oscillating boundary displacement $y(t)$ as:
\begin{equation}
\operatorname{sgn}(\phi(t,x)) =\Pi(t,x)\Sigma(t)= \frac{1}{2\pi}\sum_{m =1,3,5,\ldots}^{\infty}\frac{4}{\pi m}\sin(\omega_m t)\int_{-\infty}^{\infty}dk\, e^{ikx}\hat{\Pi}(t,k),
\end{equation}
where the spatial Fourier transform of the moving top-hat window function \eqref{eq:pi_heaviside} is given explicitly by:
\begin{equation}
\hat{\Pi}(t,k) = \int_{-\infty}^{\infty}dx\, \Pi(t,x) e^{-ikx} = e^{-iky(t)} \frac{2 \sin\left(\frac{kT}{2}\right)}{k}.
\end{equation}
To decouple the spatial differential operators, we map the complete signum-Gordon equation into wavenumber space ($x \to k$). The field equation becomes:
\begin{equation}
\partial_t^2 \hat{\phi}(t,k) + k^2 \hat{\phi}(t,k) + \widehat{\operatorname{sgn}(\phi)}(t,k) = 0.
\end{equation}
Substituting our full Fourier series ansatz for $\hat{\phi}(t,k)$ yields:
\begin{align}
\sum_{n=1}^{\infty} &\left[ (k^2 - \omega_n^2)\hat{a}_n(k)\cos(\omega_n t) + (k^2 - \omega_n^2)\hat{b}_n(k)\sin(\omega_n t) \right] \nonumber \\
&+ \frac{8\sin\left(\frac{kT}{2}\right)}{k} \sum_{m \text{ odd}}^{\infty} \frac{1}{\pi m} \sin(\omega_m t) e^{-iky(t)} = 0.
\end{align}

To isolate the spatial spectral coefficients $\hat{a}_n(k)$ and $\hat{b}_n(k)$, we exploit the piecewise kinematics of the swaying boundary by splitting the time-projection integrals into the two constituent half-periods, $t \in [0, T/2]$ and $t \in [T/2, T]$. Accounting for the standard normalization factor $T/2$ arising from the basis orthogonality over the full period, we introduce the global spectral prefactor
\begin{equation}
\mathcal{C}(k) \equiv \frac{16\sin\left(\frac{kT}{2}\right)}{\pi k T}.
\end{equation}
Crucially, while the formal boundary matching of the source envelope initially introduces an infinite series over all odd modes $m$, the physical hierarchy of the nonlinear perturbation theory dictates that higher harmonics are systematically driven by lower-order terms. Consequently, to maintain structural consistency up to the second order in the expansion of $\phi(t,x)$ and prevent the injection of unphysical high-frequency noise, the summation over $m$ is truncated at the current harmonic index $n$. As we will show later, in the static limit $v_0 \to 0$, this truncation criterion is both necessary and sufficient to capture the exact resonant contribution ($m=n$) required to properly reconstruct the unperturbed stationary core of the localized oscillon configuration. Under this truncation rule, the projection channels can be written compactly as:
\begin{align*}
\mathcal{I}_n^c(k) &= \mathcal{C}(k) \sum_{\substack{m=1 \\ m \text{ odd}}}^n \frac{1}{m} \left[ \int_{0}^{T/2} \sin(\omega_m t) \cos(\omega_n t) e^{-ikv_0 t} \, dt + \int_{T/2}^{T} \sin(\omega_m t) \cos(\omega_n t) e^{-ikv_0(T-t)} \, dt \right], \\
\mathcal{I}_n^s(k) &= \mathcal{C}(k) \sum_{\substack{m=1 \\ m \text{ odd}}}^n \frac{1}{m} \left[ \int_{0}^{T/2} \sin(\omega_m t) \sin(\omega_n t) e^{-ikv_0 t} \, dt + \int_{T/2}^{T} \sin(\omega_m t) \sin(\omega_n t) e^{-ikv_0(T-t)} \, dt \right].
\end{align*}
Using the integrals $\mathcal{I}_n^c(k)$ and $\mathcal{I}_n^s(k)$, the coefficients $\hat{a}_n(k)$ and $\hat{b}_n(k)$ can be expressed as
\[
\hat{a}_n(k)=\frac{\mathcal{I}_n^c(k)}{\omega_n^2-k^2},\qquad\qquad \hat{b}_n(k)=\frac{\mathcal{I}_n^s(k)}{\omega_n^2-k^2}.
\]
By introducing the linear change of variables $t' = T - t$ in the second half-period integrals and recalling that $\omega_j T = 2\pi j$, the shifted trigonometric factors map via $\sin(\omega_m T - \omega_m t') = -\sin(\omega_m t')$, $\cos(\omega_n T - \omega_n t') = \cos(\omega_n t')$, and $\sin(\omega_n T - \omega_n t') = -\sin(\omega_n t')$. Substituting these relations back into the second-half intervals yields:
\begin{align*}
\int_{T/2}^{T} \sin(\omega_m t) \cos(\omega_n t) e^{-ikv_0(T-t)} \, dt &= -\int_{0}^{T/2} \sin(\omega_m t') \cos(\omega_n t') e^{-ikv_0 t'} \, dt', \\
\int_{T/2}^{T} \sin(\omega_m t) \sin(\omega_n t) e^{-ikv_0(T-t)} \, dt &= +\int_{0}^{T/2} \sin(\omega_m t') \sin(\omega_n t') e^{-ikv_0 t'} \, dt'.
\end{align*}
The two half-period contributions in the cosine channel possess opposite signs and cancel identically ($\mathcal{I}_n^c(k) = 0 \implies \hat{a}_n(k) = 0$), whereas the sine channels constructively reinforce. 

Employing the product-to-sum identity $2\sin(\omega_m t)\sin(\omega_n t) = \cos(\Omega_- t) - \cos(\Omega_+ t)$ with $$\Omega_{\pm} \equiv \omega_m \pm \omega_n,$$ the factor of two from the combined half-periods cancels cleanly, leaving the total-period sine projection as:
\begin{equation}
\mathcal{I}_n^s(k) = \mathcal{C}(k)\sum_{\substack{m=1 \\ m \text{ odd}}}^n \frac{1}{m} \int_{0}^{T/2} \Big[ \cos(\Omega_- t) - \cos(\Omega_+ t) \Big] e^{-ikv_0 t}\, dt .
\end{equation}
Evaluating these elementary exponential integrals over the half-period boundaries and invoking the modal phase constraints -- namely $\sin(\frac{\Omega_{\pm} T}{2}) = 0$ and $\cos(\frac{\Omega_{\pm} T}{2}) = -(-1)^n$ for strictly odd $m$ -- yields the explicit analytical representation for the surviving driving force spectrum:
\begin{equation}\label{eq:I_s_final_condensed}
\mathcal{I}_n^s(k) =iv_0k\, \mathcal{C}(k) \left(1 + (-1)^n e^{-i k v_0 \frac{T}{2}}\right) \sum_{\substack{m=1 \\ m \text{ odd}}}^n \frac{1}{ m} \left[ \frac{1}{\Omega_-^2 - k^2 v_0^2} - \frac{1}{\Omega_+^2 - k^2 v_0^2} \right].
\end{equation}
An essential physical consequence of Eq. \eqref{eq:I_s_final_condensed} is that it yields a non-vanishing spectrum $\mathcal{I}_n^s(k)$ for even harmonic channels ($n = 2, 4, 6, \dots$) exclusively when $v_0 \neq 0$. In the strict static limit, the profile of the source is dictated by a spatial step-like configuration (a signum-type function) whose Fourier expansion possesses support solely over the odd orthogonal modes $m$. However, the introduction of a non-zero boundary velocity $v_0$ breaks the spatial reflection symmetry of the background configuration. 

Dynamic modulation couples the unperturbed odd source components to even response channels. This symmetry breaking serves as a kinematic conduit for energy transfer, dynamically dressing the moving compacton core with even-harmonic excitations that are otherwise forbidden by the parity constraints of a compact oscillon with static boundaries.

To establish the structural integrity of the dynamic tracking framework, we analyze the static limit ($v_0 \to 0$) of the total-period sine projection \eqref{eq:I_s_final_condensed}. The limit maps into two distinct physical regimes depending exclusively on the parity of the responding field channel $n$.

For strictly {\it even integers} $n$, the geometric boundary factor satisfies $(-1)^n = +1$, reducing the tracking phase parenthesis to a finite constant:
\begin{equation}
\lim_{v_0 \to 0} \left(1 + e^{-i k v_0 T / 2}\right) = 2.
\end{equation}
Because the internal frequency combinations $\Omega_{\pm} = \omega_m \pm \omega_n$ represent the coupling of an odd source $m$ and an even response $n$, the denominators $\Omega_{\pm}^2$ remain strictly non-zero for all terms in the summation. Consequently, the linear prefactor $v_0$ drives the entire channel to vanish identically:
\begin{equation}\label{eq:even_limit_vanish}
\lim_{v_0 \to 0} \mathcal{I}_n^s(k) \propto \lim_{v_0 \to 0} \left[ 2 v_0 \left( \frac{1}{\Omega_-^2} - \frac{1}{\Omega_+^2} \right) \right] = 0,\qquad n=2,4,6,\ldots
\end{equation}

For strictly {\it odd integers} $n$, the parity factor yields $(-1)^n = -1$, creating a destructive interference tracking channel. Expanding the exponential term to first order in $v_0$ isolates a linear velocity scaling:
\begin{equation}
\lim_{v_0 \to 0} \left(1 - e^{-i k v_0 T / 2}\right) \approx \frac{i k v_0 T}{2}.
\end{equation}
Substituting this expansion back into \eqref{eq:I_s_final_condensed} balances the vanishing velocity scales, yielding:
\begin{equation}
\lim_{v_0 \to 0} \mathcal{I}_n^s(k) = -8 k \sin\left(\frac{kT}{2}\right) \sum_{\substack{m=1 \\ m \text{ odd}}}^n \frac{1}{\pi m} \lim_{v_0 \to 0} \left[ \frac{v_0^2}{\Omega_-^2 - k^2 v_0^2} - \frac{v_0^2}{\Omega_+^2 - k^2 v_0^2} \right].
\end{equation}
While all non-resonant terms ($m \neq n$) vanish as $\mathcal{O}(v_0^2)$, the difference frequency vanishes identically ($\Omega_- = 0$) in the resonant channel where the source matches the response mode ($m = n$). This identity marks a physical resonance where the internal oscillation frequency of the static localized solution matches the driver frequency of the moving boundaries. This specific indeterminate form resolves to a finite constant:
\begin{equation}
\lim_{v_0 \to 0} \left[ \frac{v_0^2}{0 - k^2 v_0^2} \right] = -\frac{1}{k^2}.
\end{equation}
Mathematically, the $1/k^2$ singularity acts as a Green's function kernel for the static differential operator. Extracting this singular surviving term from the summation removes the dynamic regularization induced by the velocity, thereby recovering the exact unperturbed spectral configuration of the isolated static compacton:
\begin{equation}\label{eq:odd_limit_recover}
\lim_{v_0 \to 0}\mathcal{I}_n^s(k) = \frac{8 \sin\left(\frac{kT}{2}\right)}{\pi n k}, \qquad n=1,3,5,\ldots
\end{equation}
The dual limits \eqref{eq:even_limit_vanish} and \eqref{eq:odd_limit_recover} demonstrate the complete mathematical consistency of our dynamic tracking integration scheme, proving that the swaying boundary model preserves the underlying topological properties of the stationary core.

The explicit coordinate-space representation of the spatial profiles $b_n(x)$ is recovered by mapping the uncoupled algebraic system back via the inverse spatial Fourier transform:
\[
b_n(x) = \frac{1}{2\pi} \int_{-\infty}^{\infty} \hat{b}_n(k) e^{ikx} \, dk= \frac{1}{2\pi} \int_{-\infty}^{\infty} \frac{\mathcal{I}_n^s(k)}{\omega_n^2 - k^2} e^{ikx} \, dk
\]
which takes the form
\begin{align*}
 b_n(x)&=\frac{8i v_0}{\pi^2 T}  \sum_{\substack{m=1 \\ m \text{ odd}}}^n \frac{1}{m} \int_{-\infty}^{\infty} \frac{\sin\left(\frac{kT}{2}\right) \left(1 + (-1)^n e^{-i k v_0 \frac{T}{2}}\right) e^{ikx}}{\omega_n^2 - k^2} \left[ \frac{1}{\Omega_-^2 - k^2 v_0^2} - \frac{1}{\Omega_+^2 - k^2 v_0^2} \right] \, dk.\label{eq:inverse_fourier}
\end{align*}
To compute this efficiently using the Cauchy residue theorem  we explicitly separate the integral into two distinct sectors governed by the respective frequency poles: $b_n(x) = b_n^-(x) - b_n^+(x)$.
Using  $\sin\left(\frac{kT}{2}\right) = \frac{e^{i\frac{kT}{2}} - e^{-i\frac{kT}{2}}}{2i}$, the numerator expands into four independent spatial characteristics labeled by $j$:
\begin{equation*}
b_n^{\pm}(x) = \frac{4 v_0}{\pi^2 T}  \sum_{\substack{m=1 \\ m \text{ odd}}}^n \frac{1}{m} \sum_{j=1}^4 c_j \int_{-\infty}^{\infty} \frac{e^{ikX_j}}{(\omega_n^2 - k^2)(\Omega_{\pm}^2 - k^2 v_0^2)} \, dk
\end{equation*}
where the phase factors $c_j$ and coordinate characteristics $X_j$ are defined as:
\begin{equation}
\label{eq:characteristics_combined}
\begin{aligned}
X_1 &= x + \frac{T}{2}, &\qquad\qquad c_1 &= +1, \\
X_2 &= x - \frac{T}{2}, & c_2 &= -1, \\
X_3 &= x + \frac{T}{2} - \frac{v_0 T}{2}, & c_3 &= +(-1)^n, \\
X_4 &= x - \frac{T}{2} - \frac{v_0 T}{2}, & c_4 &= -(-1)^n.
\end{aligned}
\end{equation}

The poles lie on the real axis. Evaluating the Cauchy Principal Value (PV) via a symmetrically indented boundary contour yields:
\begin{equation}
{\rm PV} \int_{-\infty}^{\infty} \frac{e^{ikX_j}}{D_{\pm}(k)} \, dk = i\pi  \operatorname{sgn}(X_j) \sum_{s} \operatorname{Res}\left( \frac{e^{ikX_j}}{D_{\pm}(k)}, k_s \right)
\end{equation}
where $D_{\pm}(k) = (\omega_n^2 - k^2)(\Omega_{\pm}^2 - k^2 v_0^2)$. Each sub-integral possesses simple poles at $k = \pm\omega_n$ and $k = \pm\frac{\Omega_{\pm}}{v_0}$. It should be explicitly noted that the $\pm$ sign designating the independent frequency sectors ($D_+$ versus $D_-$) is completely uncorrelated with the $\pm$ sign indicating the spatial direction or orientation of the poles on the complex plane ($k = \pm \omega_n$, $k = \pm \frac{\Omega_{\pm}}{v_0}$). These two index notations operate independently and must be treated as separate degrees of freedom during the contour tracking and subsequent residue summation.

Applying the standard residue theorem formula $\operatorname{Res}(f/g, k_0) = f(k_0)/g'(k_0)$ to the simple poles of the isolated frequency sectors yields the following explicit contributions:
\begin{enumerate}
    \item \textbf{Residues at $k = \pm\omega_n$:}
    \begin{equation*}
    \operatorname{Res}\left(\frac{e^{ikX_j}}{D_{\pm}(k)}, \pm\omega_n\right)  = \mp \frac{e^{\pm i \omega_n X_j}}{2\omega_n(\Omega_{\pm}^2 - \omega_n^2 v_0^2)},
    \end{equation*}
    
    \item \textbf{Residues at $k = \pm\frac{\Omega_{\pm}}{v_0}$:}
    \begin{equation*}
    \operatorname{Res}\left(\frac{e^{ikX_j}}{D_{\pm}(k)}, \pm\frac{\Omega_{\pm}}{v_0}\right) =  \pm \frac{v_0 e^{\pm i \frac{\Omega_{\pm}}{v_0} X_j}}{2\Omega_{\pm}(\Omega_{\pm}^2 - \omega_n^2 v_0^2)}.
    \end{equation*}
\end{enumerate}

Combining the paired residues under the reflection properties of the complex exponential yields the clean, real-valued trigonometric form for each phase $X_j$:
\begin{equation}
{\rm PV} \int_{-\infty}^{\infty} \frac{e^{ikX_j}}{D_{\pm}(k)} \, dk = \pi \operatorname{sgn}(X_j) \left[ \frac{\sin(\omega_n X_j)}{\omega_n(\Omega_{\pm}^2 - \omega_n^2 v_0^2)} - \frac{v_0 \sin\left(\frac{\Omega_{\pm}}{v_0} X_j\right)}{\Omega_{\pm}(\Omega_{\pm}^2 - \omega_n^2 v_0^2)} \right]
\end{equation}

The sign modification factor $\operatorname{sgn}(X_j)$ dynamically flips depending on the observer's position along the spatial coordinate $x$. Assuming $v_0 > 0$, the roots of the characteristic equations partition the spatial domain into 5 distinct computational regions, as summarized in Table~\ref{tab:sign_distribution}. 

It is important to emphasize that changing the velocity parameter from $v_0$ to $-v_0$ is not automatic. Such an inversion corresponds to an oscillon initially moving to the left, which fundamentally alters the kinematics and requires a careful redefinition of the characteristic coordinates $X_j$. However, the solutions for the $-v_0$ case can be mapped directly from the analyzed $v_0 > 0$ case via a combined spacetime transformation. Specifically, this is achieved by applying a time shift of $T/2$ alongside a spatial shift that realigns the right boundary at $t=0$ to the position $x = T/2$.

\begin{table}[h!]
\centering
\caption{Sign distribution of $\operatorname{sgn}(X_j)$ across spacetime domains for $v_0>0$}
\label{tab:sign_distribution}
\begin{tabular}{lccccc}
\toprule
Region & Spatial Interval & $\operatorname{sgn}(X_1)$ & $\operatorname{sgn}(X_2)$ & $\operatorname{sgn}(X_3)$ & $\operatorname{sgn}(X_4)$ \\
\midrule
\textbf{I}   & $x < -\frac{T}{2}$                                     & $-1$ & $-1$ & $-1$ & $-1$ \\
\textbf{II}  & $-\frac{T}{2} < x < -(1-v_0)\frac{T}{2} $     & $+1$ & $-1$ & $-1$ & $-1$ \\
\textbf{III} & $-(1-v_0)\frac{T}{2}  < x < \frac{T}{2}$      & $+1$ & $-1$ & $+1$ & $-1$ \\
\textbf{IV}  & $\frac{T}{2} < x < (1+v_0)\frac{T}{2}$        & $+1$ & $+1$ & $+1$ & $-1$ \\
\textbf{V}   & $x > (1+v_0)\frac{T}{2} $                      & $+1$ & $+1$ & $+1$ & $+1$ \\
\bottomrule
\end{tabular}
\end{table}

Let us define the localized trigonometric kernel function $\mathcal{S}_j(x, \Omega)$ for compactness:
\begin{equation}\label{eq:S_j_definition}
\mathcal{S}_j(x, \Omega) \equiv \frac{\sin(\omega_n X_j)}{\omega_n} - \frac{v_0 \sin\left(\frac{\Omega}{v_0} X_j\right)}{\Omega}
\end{equation}
By reconstructing $b_n(x) = b_n^-(x) - b_n^+(x)$, the final exact analytical closed formula describing the oscillon field $\phi(t,x)=\sum_n b_n(x)\sin(\omega_n t)$ is written as:
\begin{equation}\label{finalsolution}
\phi(t,x)= \frac{4 v_0}{\pi T}\sum_{n=1}^{\infty} \sum_{\substack{m=1 \\ m \text{ odd}}}^n \frac{1}{m} \Bigg[ \frac{\mathcal{G}(x, \Omega_-)}{\Omega_-^2 - v_0^2 \omega_n^2} - \frac{\mathcal{G}(x, \Omega_+)}{\Omega_+^2 - v_0^2 \omega_n^2} \Bigg]\sin(\omega_n t),
\end{equation}
where the generalized spatial phase modulator for a given frequency channel $\Omega \in \{\Omega_-, \Omega_+\}$ is defined as:
\begin{equation}\label{eq:G_definition}
\mathcal{G}(x, \Omega) \equiv \sum_{j=1}^4 c_j \operatorname{sgn}(X_j) \mathcal{S}_j(x, \Omega).
\end{equation}
The underlying kinematic basis functions are given by \eqref{eq:S_j_definition}, the phase factors $c_j$ and coordinate characteristics $X_j$ are defined in \eqref{eq:characteristics_combined}.

\subsection{Compactness: Vanishing of the solution in the outer region}

To establish the local finiteness and compact support of the spatial profile within our general analysis of the solution, we must demonstrate that the field configuration $b_n(x)$ vanishes identically outside the localized interaction zone. 
We begin by examining the behavior of $\mathcal{G}(x, \Omega)$ within the external asymptotic regions $\textbf{I}$ and $\textbf{V}$ before turning to the solution inside the boundary horizons.

For all spatial coordinates located in the far-left exterior domain (Region $\textbf{I}$: $x < -\frac{T}{2}$), the arguments of the signum functions are strictly negative, such that $X_j < 0$ for all $j \in \{1, 2, 3, 4\}$. This yields a uniform sign distribution $\operatorname{sgn}(X_j) = -1$. Substituting the explicit phase weights $c_1 = 1$, $c_2 = -1$, $c_3 = (-1)^n$, and $c_4 = -(-1)^n$, the total phase modulator for a fixed nonlinear frequency channel $\Omega \in \{\Omega_-, \Omega_+\}$ defined in \eqref{eq:G_definition} reduces in Region $\textbf{I}$ to:
\begin{equation}\label{eq:G_region_I}
\mathcal{G}_{\text{I}}(x, \Omega) = -\mathcal{S}_1(x, \Omega) + \mathcal{S}_2(x, \Omega) - (-1)^n \mathcal{S}_3(x, \Omega) + (-1)^n \mathcal{S}_4(x, \Omega),
\end{equation}
where the factors $\mathcal{S}_j(x, \Omega)$ are given by \eqref{eq:S_j_definition}. The total modulator decomposes into  $\mathcal{G}_{\text{I}}^{(\omega)}$ and  $\mathcal{G}_{\text{I}}^{(\Omega)}$ as $\mathcal{G}_{\text{I}}(x, \Omega) = \mathcal{G}_{\text{I}}^{(\omega)}(x) + \mathcal{G}_{\text{I}}^{(\Omega)}(x)$, which expand as:
\begin{align}
\omega_n \mathcal{G}_{\text{I}}^{(\omega)}(x) &= -\sin(\omega_n X_1) + \sin(\omega_n X_2) - (-1)^n \sin(\omega_n X_3) + (-1)^n \sin(\omega_n X_4), \label{eq:G_omega_I} \\
\frac{\Omega}{v_0} \mathcal{G}_{\text{I}}^{(\Omega)}(x) &= \sin\left(\frac{\Omega}{v_0} X_1\right) - \sin\left(\frac{\Omega}{v_0} X_2\right) + (-1)^n \sin\left(\frac{\Omega}{v_0} X_3\right) - (-1)^n \sin\left(\frac{\Omega}{v_0} X_4\right). \label{eq:G_Omega_I}
\end{align}
By employing the spatial translations matching the full swaying period, $X_2 = X_1 - T$ and $X_4 = X_3 - T$, alongside the boundary periodicity condition $\omega_n T = 2\pi n$ ($n \in \mathbb{Z}$), the arguments in \eqref{eq:G_omega_I} map via the identities $\sin(\omega_n X_2) = \sin(\omega_n X_1 - 2\pi n) = \sin(\omega_n X_1)$ and $\sin(\omega_n X_4) = \sin(\omega_n X_3)$, yielding an immediate pairwise cancellation such that $\mathcal{G}_{\text{I}}^{(\omega)}(x) \equiv 0$. To evaluate the expression \eqref{eq:G_Omega_I}, we express the characteristics $X_3$ and $X_4$ explicitly in terms of $X_1$ and $X_2$ via $X_3 = X_1 - \frac{v_0 T}{2}$ and $X_4 = X_2 - \frac{v_0 T}{2}$. Substituting these relations into the last two terms of Eq.~\eqref{eq:G_Omega_I} yields:
\begin{equation}\label{eq:kinematic_phase_shift}
\sin\left(\frac{\Omega}{v_0} X_{3,4}\right) = \sin\left(\frac{\Omega}{v_0} X_{1,2} - \frac{\Omega T}{2}\right) = -(-1)^n \sin\left(\frac{\Omega}{v_0} X_{1,2}\right).
\end{equation}
Here, we have exploited the discrete spectral relation $\frac{\Omega T}{2} = \pi(m \pm n)$ under the structural constraint that $m$ runs exclusively over odd harmonic integers ($m = 2k+1, \, k \in \mathbb{Z}$), which dictates $\cos\left(\frac{\Omega T}{2}\right) = -(-1)^n$. Inserting the identity \eqref{eq:kinematic_phase_shift} back into Eq.~\eqref{eq:G_Omega_I} ensures a simultaneous algebraic cancellation across all terms, proving that $\mathcal{G}_{\text{I}}^{(\Omega)}(x) \equiv 0$, and thus the total left exterior modulator vanishes identically.

An identical structural mechanism manifests on the opposite side of the configuration. In the far-right external domain (Region $\textbf{V}$: $x > \frac{T}{2} + \frac{v_0 T}{2}$), all coordinate characteristics transcend the moving wavefront boundaries ($X_j > 0$), producing a uniform positive sign distribution $\operatorname{sgn}(X_j) = +1$. The total phase modulator in Region $\textbf{V}$ mirrors the left domain up to a global sign reflection, $\mathcal{G}_{\text{V}}(x, \Omega) = +\mathcal{S}_1(x, \Omega) - \mathcal{S}_2(x, \Omega) + (-1)^n \mathcal{S}_3(x, \Omega) - (-1)^n \mathcal{S}_4(x, \Omega) = -\mathcal{G}_{\text{I}}(x, \Omega)$. Applying the exact same spatial translations and kinematic shifts, the components within $\mathcal{G}_{\text{V}}(x, \Omega)$ cancel out entirely, showing that $\mathcal{G}_{\text{V}}(x, \Omega) \equiv 0$. The simultaneous vanishing of the modulators in both external limits yields the global spatial identity $b_n(x) \equiv 0$ for all $x \in \left(-\infty, -\frac{T}{2}\right) \cup \left(\frac{T}{2} + \frac{v_0 T}{2}, \infty\right)$. This rigorous demonstration confirms that the radiative components traveling at disparate phase velocities undergo complete destructive interference outside the moving horizons, establishing the compact support of the dynamic oscillon configuration for any arbitrary boundary velocity $v_0 \in (-1,1)$.

\subsection{Inner region: Resolution of indeterminate forms}
With the zero-field conditions firmly established for the exterior domains, we can now proceed to track the detailed internal profile spanning Regions $\textbf{II}$, $\textbf{III}$, and $\textbf{IV}$.

The solution \eqref{finalsolution} describing the compact oscillon may superficially look poorly behaved for certain combinations of parameters and the integers labeling the oscillon's Fourier modes. We will show here that all indeterminate forms of the type $\frac{0}{0}$ pose no physical problem and that the solution behaves correctly at these specific points.

Analyzing the behavior of the expression \eqref{eq:S_j_definition}, we see from the definition $\Omega_{\pm}=\frac{2\pi}{T}(m\pm n)$ that the combined frequency $\Omega_-$ vanishes for $m=n$ (with both $m$ and $n$ being odd integers). Since the second term in the definition of $\mathcal{S}_j(x, \Omega_-)$ contains $\Omega_-$ in both the argument of the sine function and the denominator, the expression takes an indeterminate form of the type $\frac{0}{0}$. Evaluating the limit as $\Omega_- \to 0$ via l'H\^{o}pital's rule yields the regular result:
\begin{equation}\label{eq:S_j_limit_Omega_minus}
\lim_{\Omega_- \to 0} \mathcal{S}_j(x, \Omega_-) = \frac{\sin(\omega_n X_j)}{\omega_n} - X_j.
\end{equation}

Another indeterminate form appears when analyzing the ratio $\frac{\mathcal{G}(x, \Omega_-)}{\Omega_-^2 - v_0^2 \omega_n^2}$ in \eqref{finalsolution}. According to \eqref{eq:G_definition}, this expression is essentially a linear combination of individual blocks of the form $\frac{\mathcal{S}_j(x, \Omega_-)}{\Omega_-^2 - v_0^2 \omega_n^2}$, which reduce to indeterminate forms of the type $\frac{0}{0}$ for certain characteristic velocities $v_0 = v_c$. Analyzing the denominator, we see that it vanishes for
\begin{equation}\label{eq:v_critical}
v_c = \mp \frac{\Omega_-}{\omega_n} = \pm \left(1- \frac{m}{n}  \right).
\end{equation}
On the other hand, the spatial basis function satisfies $\mathcal{S}_j(x, \Omega_-) = 0$ at this specific velocity parameter $v_c$. To establish the finite limit of the ratio, we apply l'H\^{o}pital's rule by differentiating both the numerator and the denominator with respect to $v_0$, taking into account the implicit dependence of the moving boundaries $X_j(v_0)$ for $j \in \{3,4\}$. This procedure yields the regularized resonant profile:
\begin{equation}\label{eq:lhopital_result}
\lim_{v_0 \to v_c} \frac{\mathcal{S}_j(x, \Omega_-)}{\Omega_-^2 - v_0^2 \omega_n^2} = \frac{1}{2 \Omega_-^2} \left[ \frac{\sin(\omega_n X_j^c)}{\omega_n} - X_j^c \cos(\omega_n X_j^c) \right],
\end{equation}
where $X_j^c \equiv X_j(v_c)$ denotes the critical value of the coordinate characteristics evaluated at $v_0 = v_c$.


\subsection{The static boundary limit: $v_0 \to 0$ via resonance resolution}

To demonstrate the physical and mathematical consistency of the exact dynamical solution \eqref{finalsolution}, we investigate the static boundary limit where the envelope expansion velocity vanishes, $v_0 \to 0$. In this limit, the system must smoothly recover the established field configuration of a compact oscillon confined within fixed, static boundaries, given by the Fourier series:
\begin{equation}\label{eq:static_oscillon}
\phi_{\text{static}}(t,x) = \sum_{\substack{n=1 \\ n \text{ odd}}}^{\infty} \frac{4}{\pi n \omega_n^2} \Big[1 + \cos\left(\omega_n x\right) \Big]\Pi(x) \sin\left(\omega_n t\right),
\end{equation}
where the even harmonic channels vanish identically, and only the odd modes contribute to the localized wave packet.

At first glance, the presence of $v_0$ as a global linear prefactor in front of the summation in \eqref{finalsolution} might suggest that the entire dynamic field trivially vanishes as $v_0 \to 0$. However, a rigorous evaluation reveals a structural resonance within the summation over the internal source modes $m$. For a fixed odd harmonic channel $n$, a singular behavior emerges at the exact matching mode $m = n$, where the combined frequency parameter vanishes identically:
\begin{equation}
\Omega_- \equiv \omega_m - \omega_n = 0 \quad (\text{for } m = n).
\end{equation}
For all other modes where $m \neq n$, as well as for all sum-frequency channels $\Omega_+ = \omega_m + \omega_n$, the frequency combinations remain strictly finite and non-zero. Consequently, as $v_0 \to 0$, the global linear prefactor $v_0$ directly suppresses all non-resonant channels ($\mathcal{O}(v_0^2)$ in the numerator against a finite denominator). The entire infinite series over $m$ thus collapses strictly to the single resonant term $m = n$.

To evaluate this resonant contribution without ambiguity, the formal limit $\Omega_- \to 0$ must be performed prior to taking the static boundary limit $v_0 \to 0$. Let us examine the dynamic components of the kinematic spatial basis functions $\mathcal{S}_j(x, \Omega_-)$ associated with the regularized resonant channel. Applying the standard fundamental limit $\lim_{z \to 0} \frac{\sin z}{z} = 1$ directly yields:
\begin{equation}\label{eq:S_j_resonance_limit}
\lim_{\Omega_- \to 0} \left[ - \frac{v_0 \sin\left(\frac{\Omega_-}{v_0} X_j\right)}{\Omega_-} \right] = - v_0 \frac{X_j}{v_0} = - X_j.
\end{equation}
This cancellation removes the trigonometric phase oscillation, leaving a linear dependence on the coordinate characteristic  $X_j$. Concurrently, evaluating the resonant condition $\Omega_- = 0$ within the associated frequency denominators of the full amplitude formula yields 
\begin{equation}\label{eq:denominator_v0_square}
\left. \left( \Omega_-^2 - v_0^2 \omega_n^2 \right) \right|_{\Omega_- = 0} = - v_0^2 \omega_n^2,
\end{equation}
which explicitly introduces a $v_0^2$ factor in the denominator of the resonant field distribution.

Next, we collect the corresponding spatial contributions from the generalized phase modulator \eqref{eq:G_definition} for the odd $n$ harmonic channels, where the structural coupling coefficients are given by $c_1=1, c_2=-1, c_3=-1, c_4=1$. Factoring out the dynamic linear coordinate contributions obtained in \eqref{eq:S_j_resonance_limit}, the remaining linear combination of characteristics partitions itself into distinct components:
\begin{equation}\label{Gres}
\mathcal{G}_{\text{res}}(x, \Omega_- \to 0) = \frac{1}{\omega_n}\mathcal{G}_{\text{trig}} - \mathcal{G}_{\text{lin}},
\end{equation}
where the trigonometric and linear configurations are defined respectively as:
\begin{align*}
\mathcal{G}_{\text{trig}} &= \sin(\omega_n X_1) + \sin(\omega_n X_2) - \sin(\omega_n X_3) - \sin(\omega_n X_4), \\
\mathcal{G}_{\text{lin}} &= X_1 + X_2 - X_3 - X_4.
\end{align*}
Here, we have explicitly utilized the fact that within the compacton core domain $x \in [-T/2, T/2]$, the right-channel characteristics map into strictly positive coordinates ($\operatorname{sgn}(X_{1,3}) = +1$), whereas the left-channel characteristics map into strictly negative coordinates ($\operatorname{sgn}(X_{2,4}) = -1$). This precise sign distribution enforces the algebraic reductions that neutralize the discontinuous step functions at the interfaces.

We now substitute the explicit definitions of the moving boundary characteristics, which expand linearly with respect to the velocity as $X_3 = X_1^0 - \frac{v_0 T}{2}$ and $X_4 = X_2^0 - \frac{v_0 T}{2}$, while the laboratory frames remain static ($X_1 = X_1^0 = x + T/2$ and $X_2 = X_2^0 = x - T/2$). For the linear part, this directly yields:
\begin{equation}\label{Glin}
\mathcal{G}_{\text{lin}} = X_1^0 + X_2^0 - \left(X_1^0 - \frac{v_0 T}{2}\right) - \left(X_2^0 - \frac{v_0 T}{2}\right) = v_0 T.
\end{equation}
For the trigonometric part, expanding the moving terms $\sin(\omega_n X_{3,4})$ via a first-order Taylor series around $v_0 = 0$ leads to:
\begin{align*}
\mathcal{G}_{\text{trig}} &= \sin(\omega_n X_1^0) + \sin(\omega_n X_2^0) - \left[ \sin(\omega_n X_1^0) - v_0 \frac{\omega_n T}{2} \cos(\omega_n X_1^0) \right]\nonumber\\ &- \left[ \sin(\omega_n X_2^0) - v_0 \frac{\omega_n T}{2} \cos(\omega_n X_2^0) \right] + \mathcal{O}(v_0^2) \nonumber \\
&= v_0 \frac{\omega_n T}{2} \left[ \cos\left(\omega_n\left(x+\frac{T}{2}\right)\right) + \cos\left(\omega_n\left(x-\frac{T}{2}\right)\right) \right] + \mathcal{O}(v_0^2)\\
&= v_0\omega_n T \cos(\omega_n x) \cos(n\pi)+ \mathcal{O}(v_0^2).
\end{align*}
Given that $n$ is an odd integer, $\cos(n\pi) = -1$, which reduces the trigonometric modulator contribution to:
\begin{equation}\label{Gtrig}
\mathcal{G}_{\text{trig}} = -v_0 \omega_n T \cos(\omega_n x) + \mathcal{O}(v_0^2).
\end{equation}
Combining both components \eqref{Gtrig} and \eqref{Glin}, the total resonant spatial phase modulator \eqref{Gres} is expressed up to first order as:
\begin{equation}\label{eq:G_res_final_expansion}
\mathcal{G}_{\text{res}}  = -v_0 T \Big[ 1 + \cos(\omega_n x) \Big] + \mathcal{O}(v_0^2).
\end{equation}
This geometric shifting of the moving boundaries provides the necessary second factor of $v_0$ in the numerator.

Finally, inserting the resolved modulator \eqref{eq:G_res_final_expansion} and the resonant denominator \eqref{eq:denominator_v0_square} back into the global field expression \eqref{finalsolution} for the single contributing resonant mode $m = n$, we obtain the following regularized limit for the Fourier amplitude:
\begin{align*}
B_n(x) = \lim_{v_0 \to 0} \left\{ \frac{4 v_0^2 T}{\pi T n} \cdot \frac{1 + \cos(\omega_n x)}{v_0^2 \omega_n^2} \right\}.
\end{align*}
The $v_0^2$ term in the numerator cancels the $v_0^2$ term generated by the resonant denominator. Simplifying the remaining global scaling constants directly yields the invariant amplitude:
\begin{equation}
B_n(x) = \frac{4}{\pi n \omega_n^2} \Big[ 1 + \cos(\omega_n x) \Big].
\end{equation}
This formal cancellation eliminates the apparent velocity singularity and proves that the dynamic configurations converge  to the stationary compact oscillon solution.

\subsection{Graphical representation of solutions}

\begin{figure}[h!]
\subfigure[]{\includegraphics[width=0.32\linewidth]{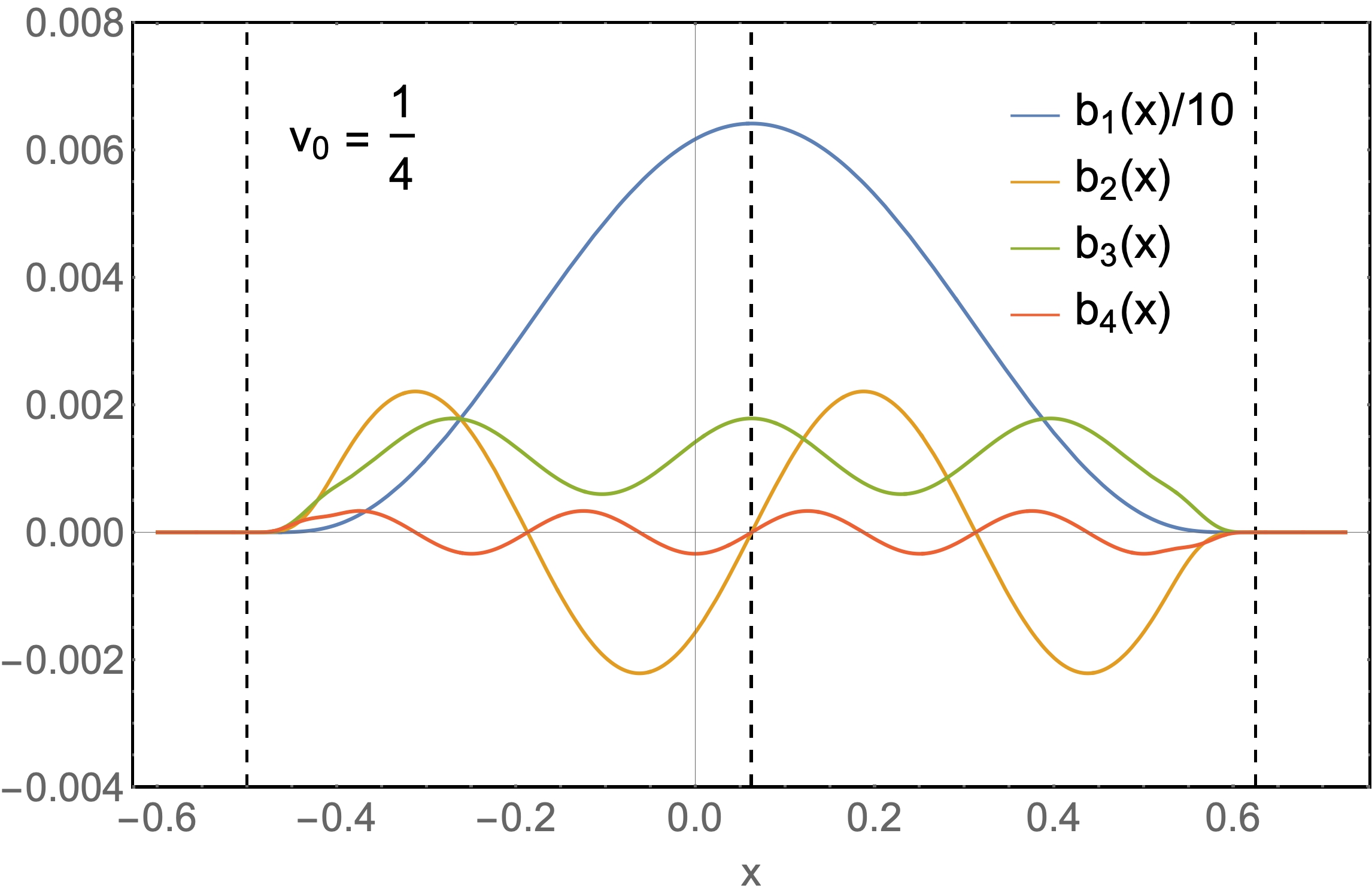}\label{fig:bn025}}
\subfigure[]{\includegraphics[width=0.32\linewidth]{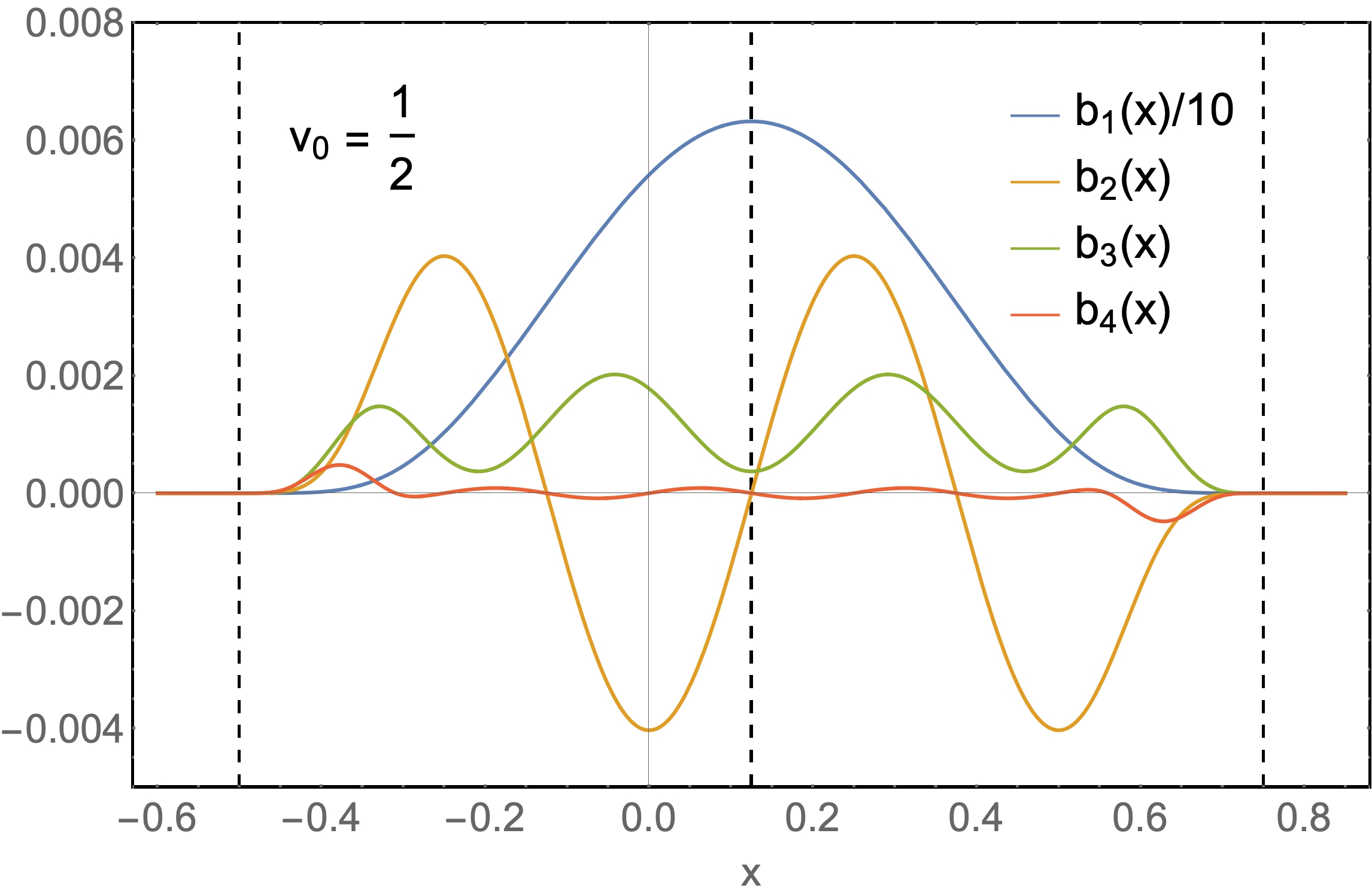}\label{fig:bn05}}
\subfigure[]{\includegraphics[width=0.32\linewidth]{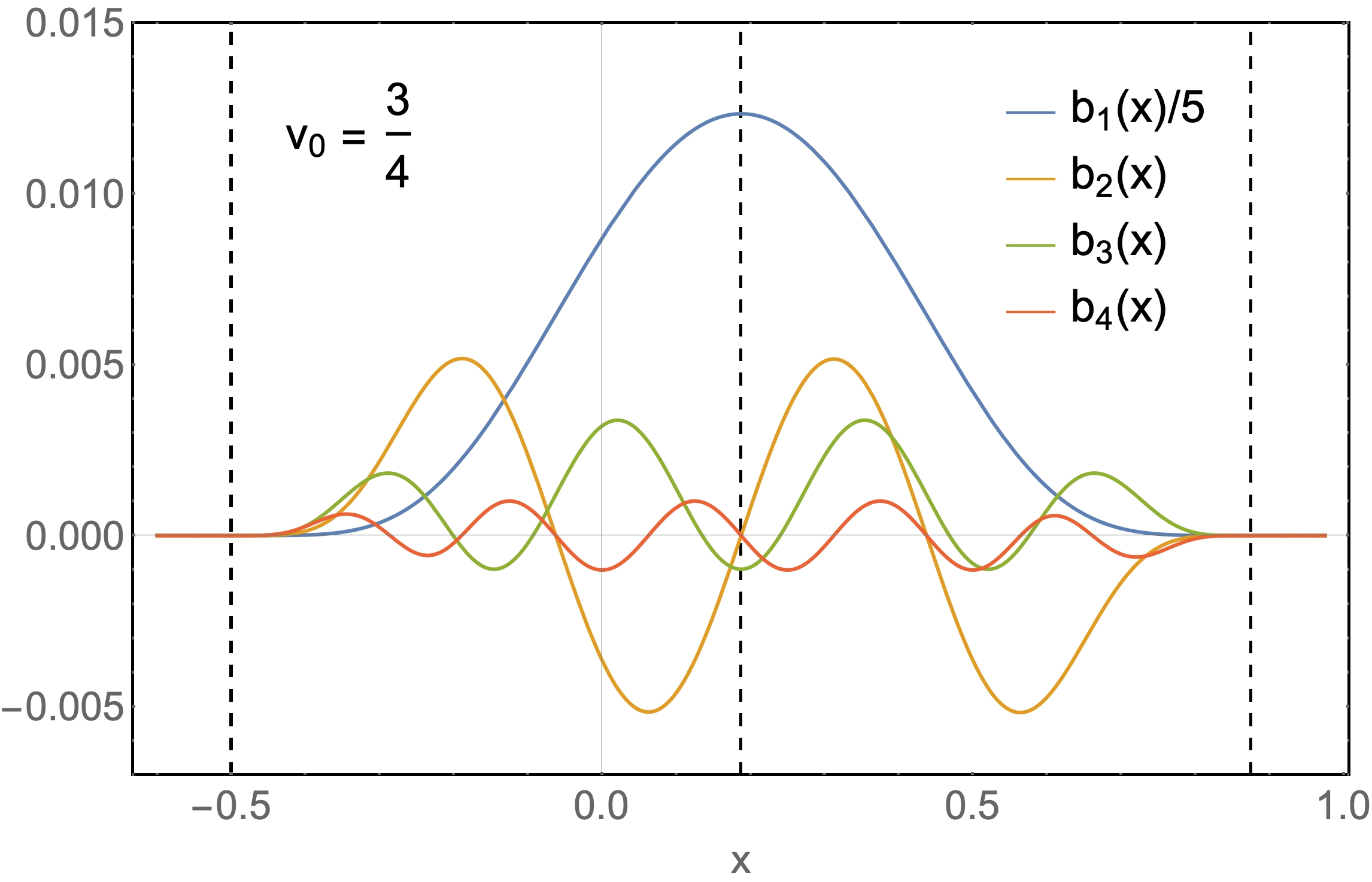}\label{fig:bn075}}
\caption{Spatial profiles of the first four expansion coefficients $b_n(x)$ for a swaying oscillon mode ($T=1$) at (a) $v_0=0.25$, (b) $v_0=0.5$, and (c) $v_0=0.75$. For enhanced visibility, $b_1(x)$ is rescaled by a factor of 10 in panels (a) and (b), and 5 in panel (c). Vertical dashed lines indicate the center and boundaries of the support interval for $b_n(x)$.} \label{fig:coefficients}
\end{figure}

We examine the truncated Fourier series representing the solutions of a swaying oscillon parameterized by the velocity $v_0 > 0$. Figure~\ref{fig:coefficients} displays the first four coefficients $b_n(x)$ for $v_0 = 0.25$, $0.5$, and $0.75$. The functions $b_n(x)$ vanish identically outside the interval $x \in [-\frac{T}{2}, (1+v_0)\frac{T}{2}]$. As $v_0$ increases, $b_3(x)$ develops additional zeros within its support--a behavior also exhibited by higher odd-order coefficients (omitted for clarity). Similarly to the $v_0=0$ case, shown in Fig.~\ref{fig:bn0}, the coefficients $b_n(x)$ for odd $n$ are symmetric with respect to the midpoint $x_c=v_0\frac{T}{4}$, whereas the even-$n$ coefficients are antisymmetric. The maximum of $b_1(x)$ occurs at $x_c$, with the value
\[
b_1(x_c)=\frac{T^2}{\pi^3}\Bigg(1+\frac{2 \sin(\frac{v_0\pi}{2})}{v_0\pi}\frac{\cos\big((1-\frac{1}{T})\frac{v_0\pi}{2}\big)}{1-\frac{v_0^2}{4}}\Bigg).
\]

\begin{figure}[h!]
\subfigure[]{\includegraphics[width=0.45\linewidth]{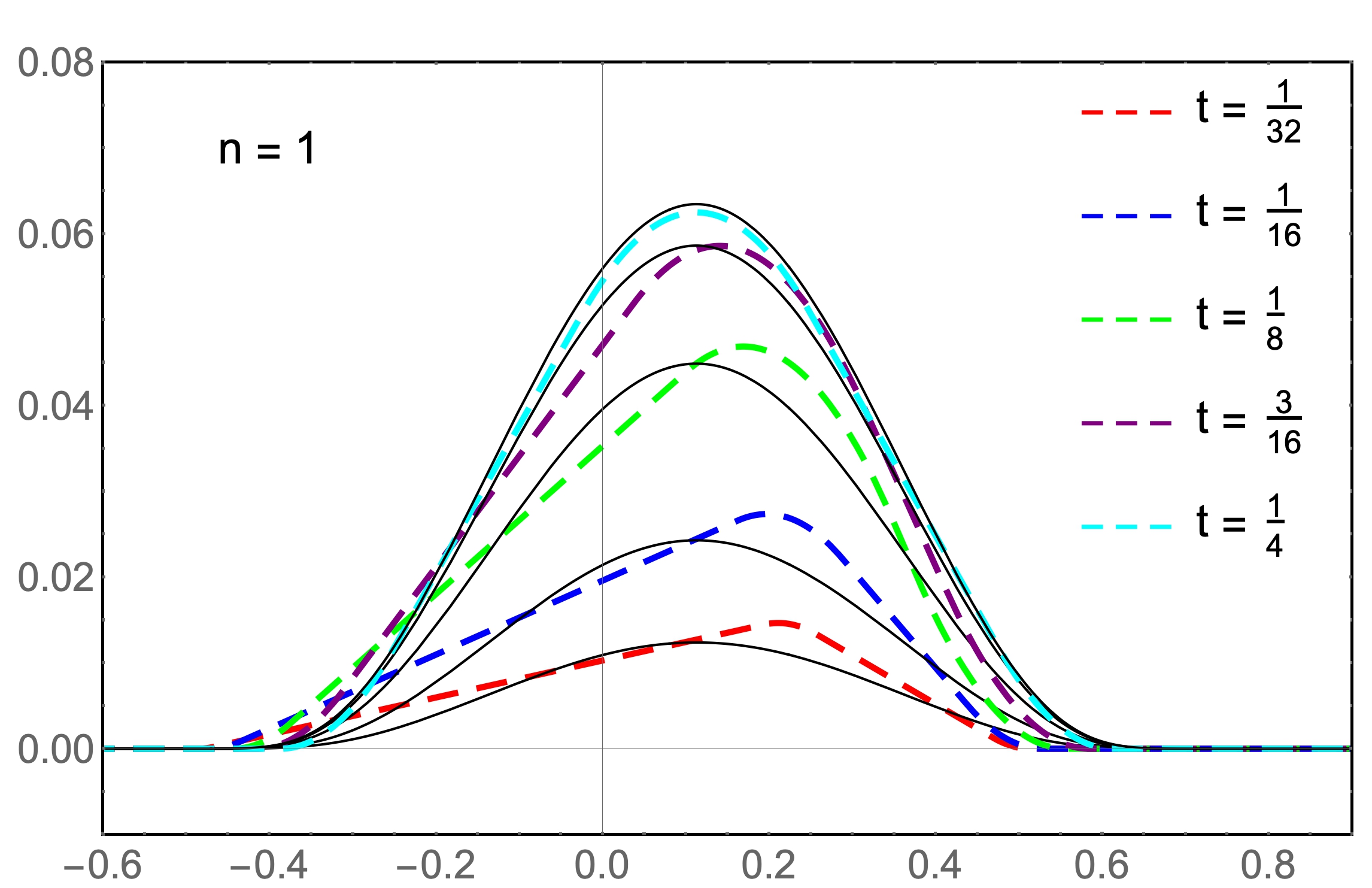}\label{fig:swp1}}
\subfigure[]{\includegraphics[width=0.45\linewidth]{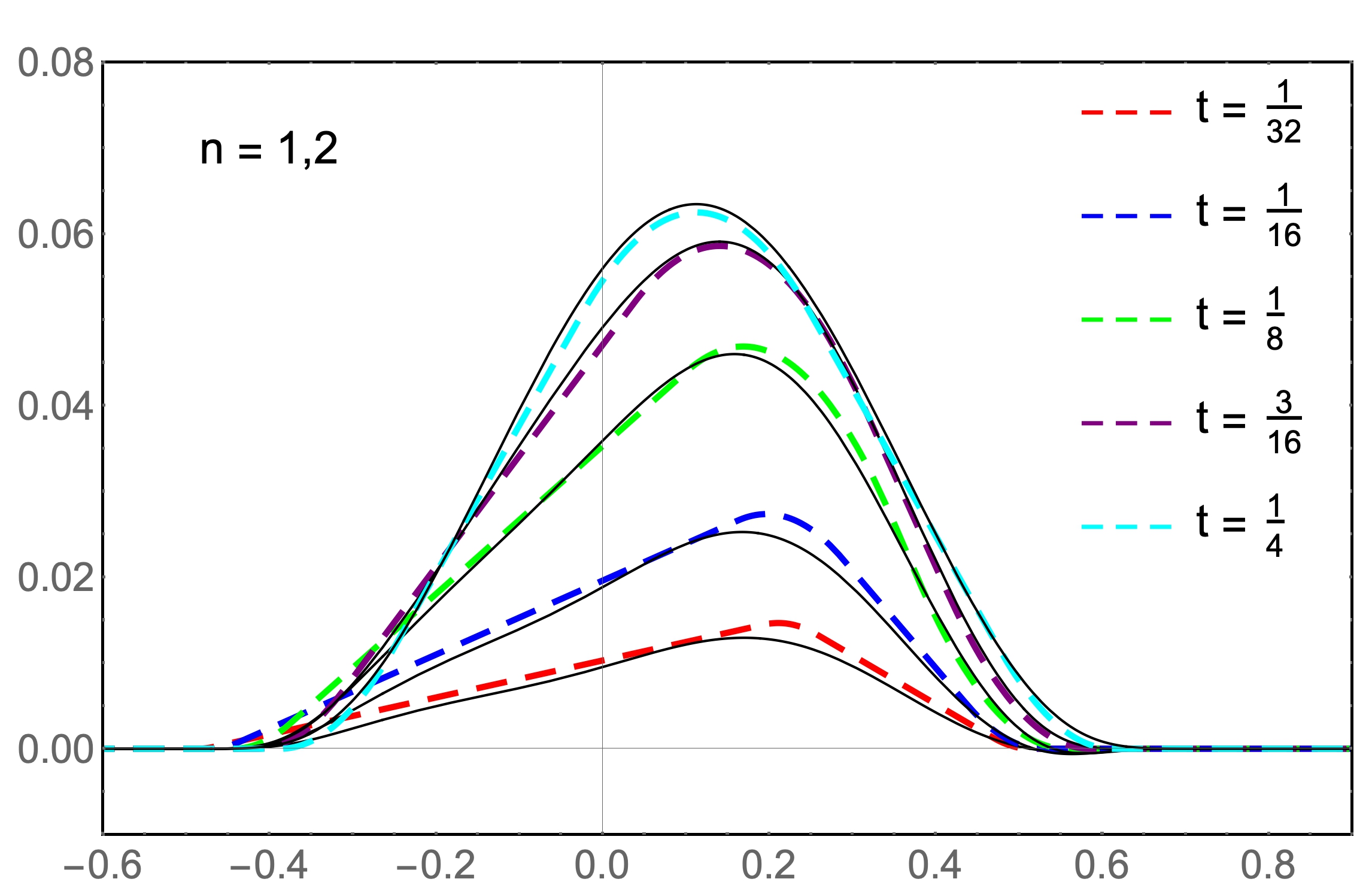}\label{fig:swp2}}
\subfigure[]{\includegraphics[width=0.45\linewidth]{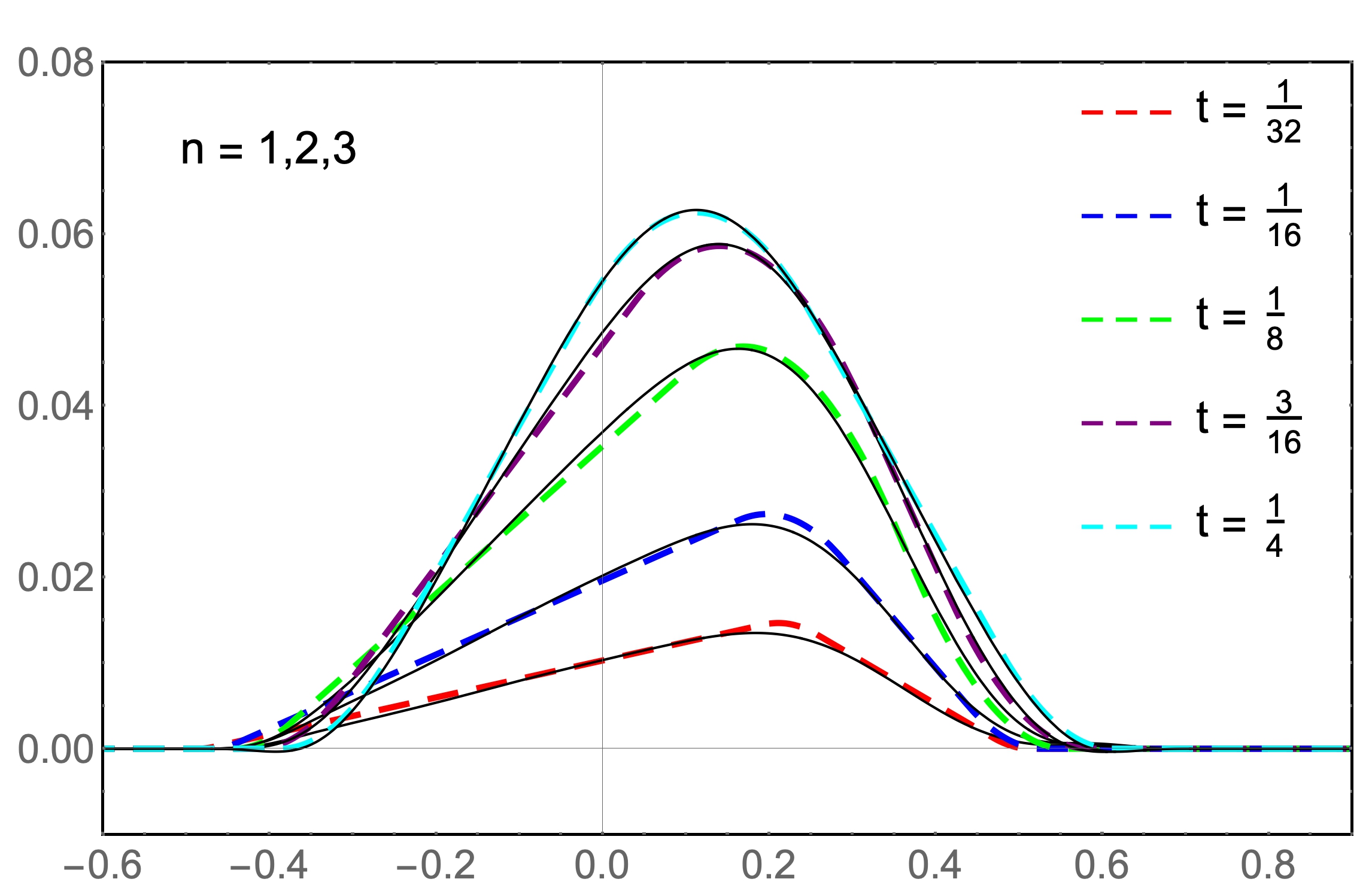}\label{fig:swp3}}
\subfigure[]{\includegraphics[width=0.45\linewidth]{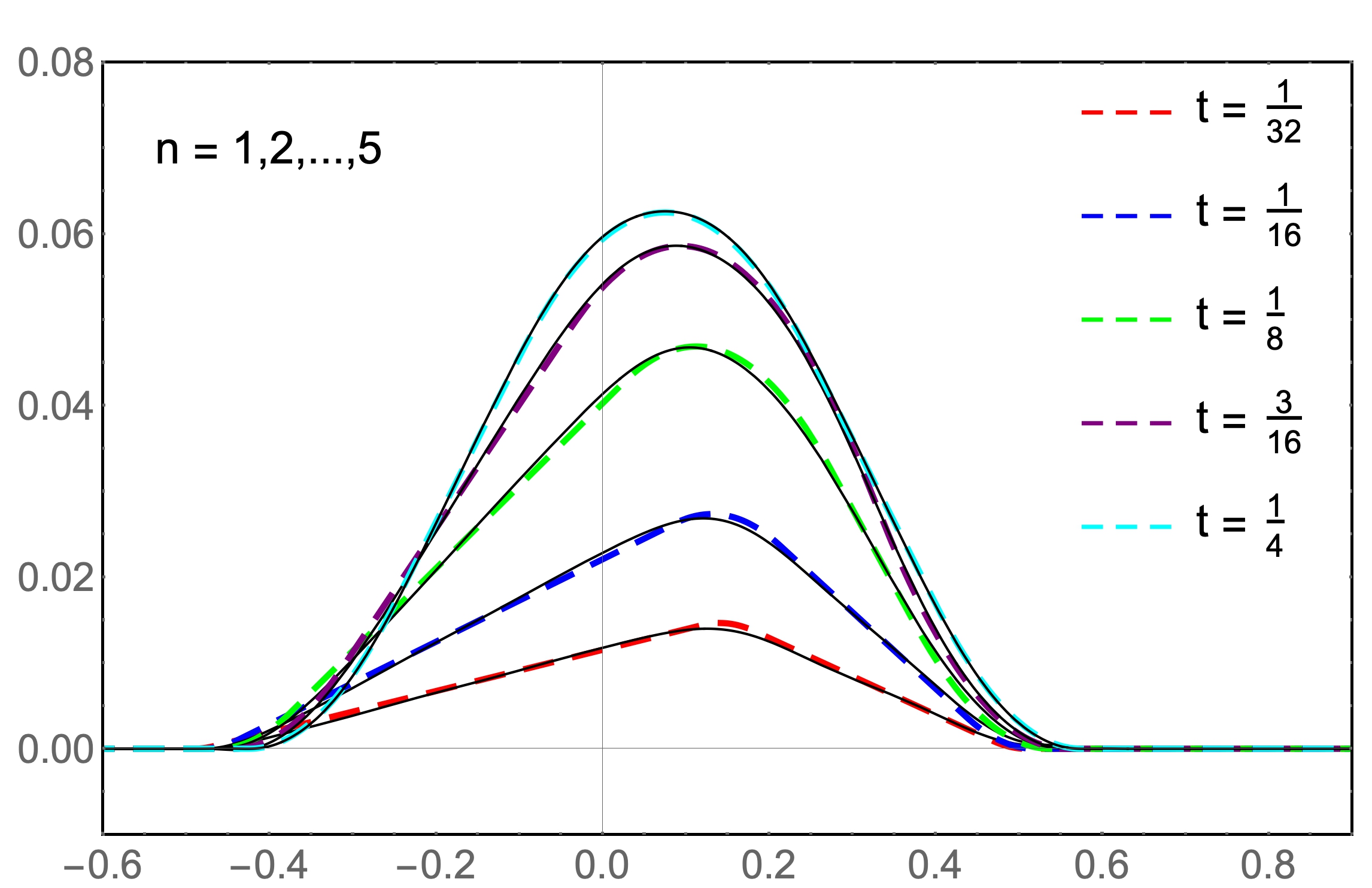}\label{fig:swp5b}}
\caption{Snapshots of the exact swaying oscillon with $v_0=0.45$ (dashed line) and its Fourier approximations containing terms with (a) $n=1$, (b) $n=1,2$, and (c) $n=1,2,3$ and (d) $n=1,2,3,4,5$.}\label{swaying-exact-fourier}
\end{figure}

Figure~\ref{swaying-exact-fourier} compares these truncated Fourier reconstructions with the exact solutions $\phi(t,x)$. For clarity, the time domain is restricted to $0 \le t \le T/4$.

When considering only the single-mode contribution ($n=1$), as shown in Fig.~\ref{fig:swp1}, the displacement of the oscillon's maximum is directly associated with the translation of the entire spatial profile. Because there are no contributions from the even channels, this lowest-order truncated solution retains a perfectly symmetric, bell-shaped curve. 

The activation of the lowest mode from the even channel ($n=2$) breaks the spatial symmetry of the oscillon shape, as illustrated in Fig.~\ref{fig:swp2}. The subsequent contributions from higher modes are depicted in Figs.~\ref{fig:swp3} and \ref{fig:swp5b}, respectively. Notably, the convergence of the truncated series to the exact solution is significantly better at instants of time when the oscillon profile is closer to being symmetric.
 
\begin{figure}[h!]
\subfigure[]{\includegraphics[width=0.32\linewidth]{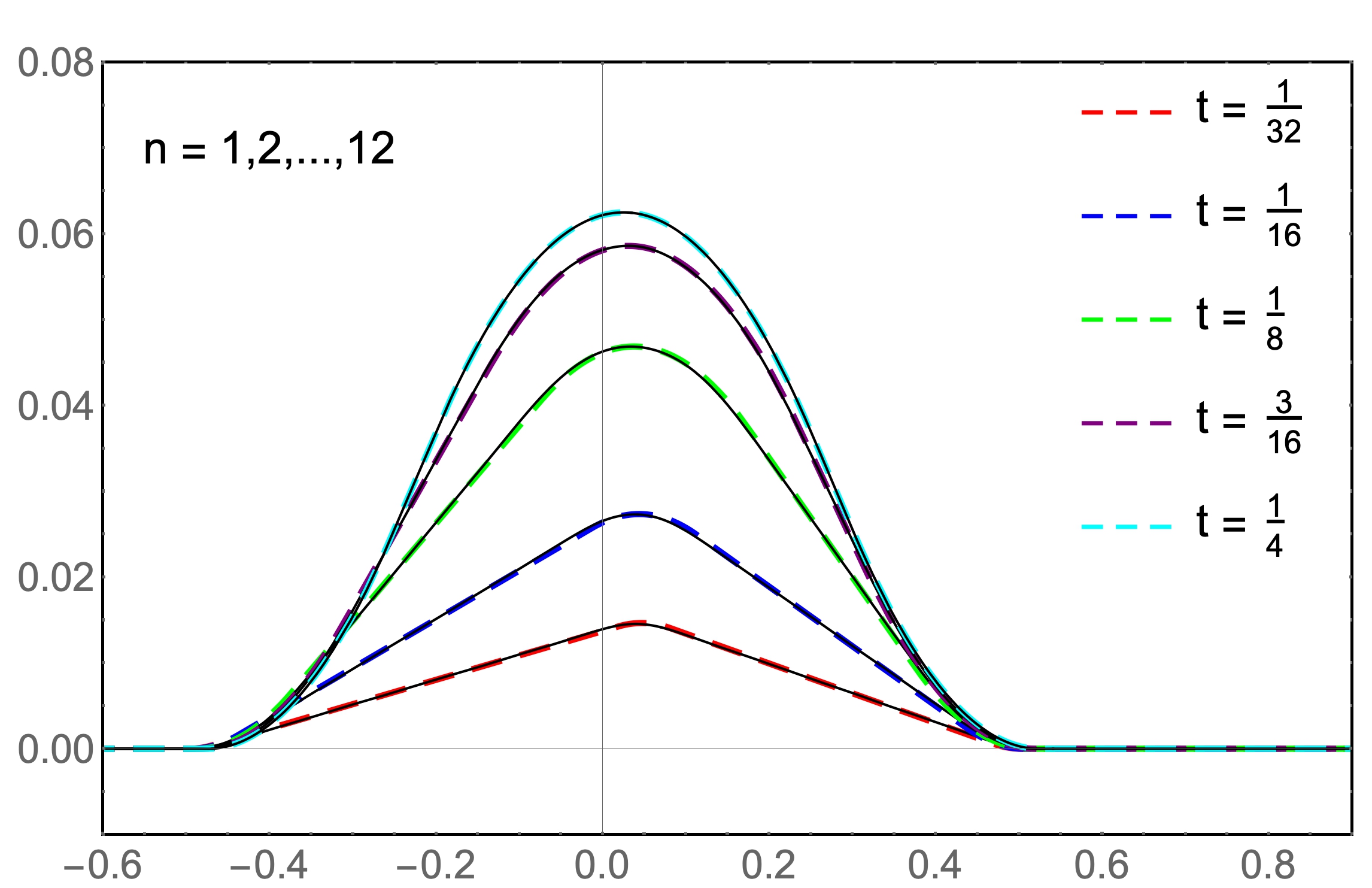}\label{fig:swp5}}
\subfigure[]{\includegraphics[width=0.32\linewidth]{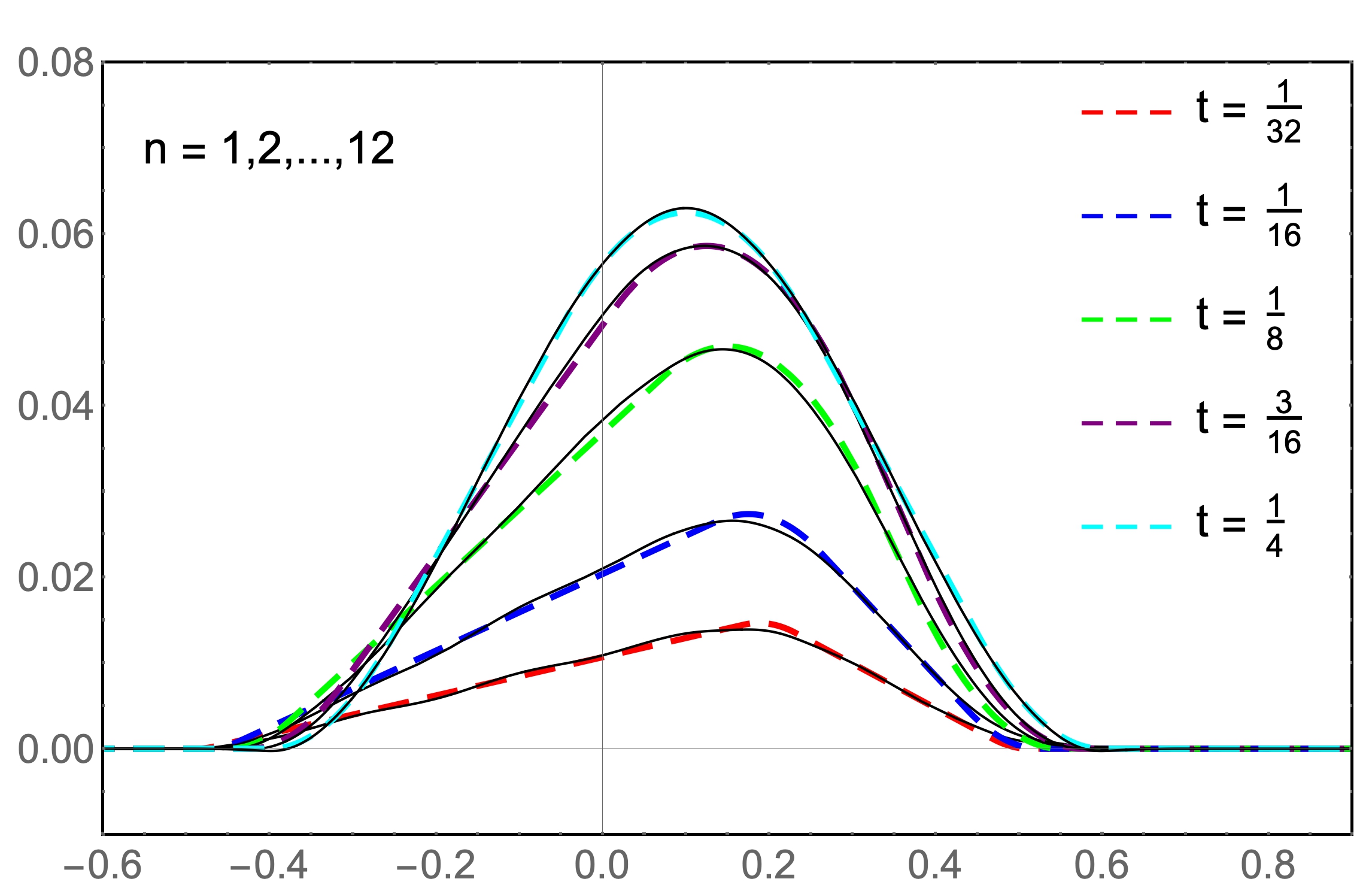}\label{fig:swp6}}
\subfigure[]{\includegraphics[width=0.32\linewidth]{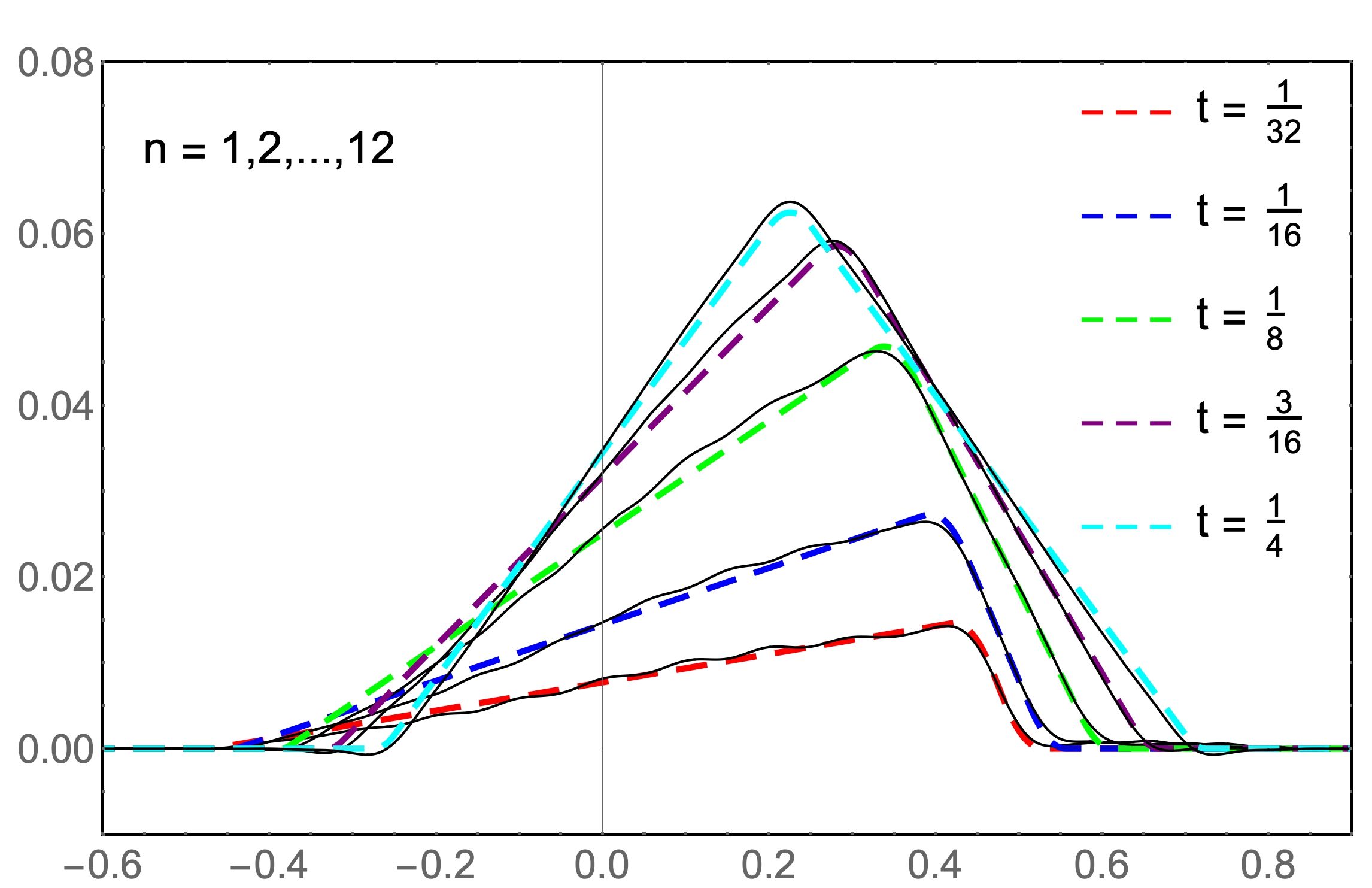}\label{fig:swp7}}
\caption{Snapshots of the exact swaying oscillon  (dashed line) and its Fourier approximations truncated to $n=12$ in dependence on  (a) $v_0=0.1$, (b) $v_0=0.3$, and (c) $v_0=0.9$.}\label{swaying-exact-fourier-velocity}
\end{figure}

Since the degree of asymmetry can be regulated by the value of $v_0$, we also investigate the convergence of the series by fixing the cutoff parameter at $n=12$ and gradually increasing the value of $v_0$. As can be seen from Fig.~\ref{swaying-exact-fourier-velocity}, the convergence deteriorates for higher values of $v_0$ compared to cases with smaller velocity parameters, reflecting a stronger excitation of non-linear modes. This problem is closely related to the Gibbs phenomenon, as in the limit $v_0 \to 1$, the oscillon shape tends toward a triangular form, with the emergence of a sharp edge at the oscillon maximum. Consequently, the first derivative of the profile becomes discontinuous at this peak. Mathematically, this transition from a smooth, analytic profile to a piecewise smooth one drastically slows the asymptotic decay rate of the Fourier coefficients. As a result, Fourier convergence is faster for smoother oscillon profiles than for sharper ones in the limit $\vert{}v_0\vert{} \to 1$. This slow decay forces truncated Fourier series to exhibit characteristic high-frequency oscillations and an unavoidable overshoot near the maximum, explaining the poor convergence observed in the high-velocity limit.

\begin{figure}[h!]
{\includegraphics[width=0.9\linewidth]{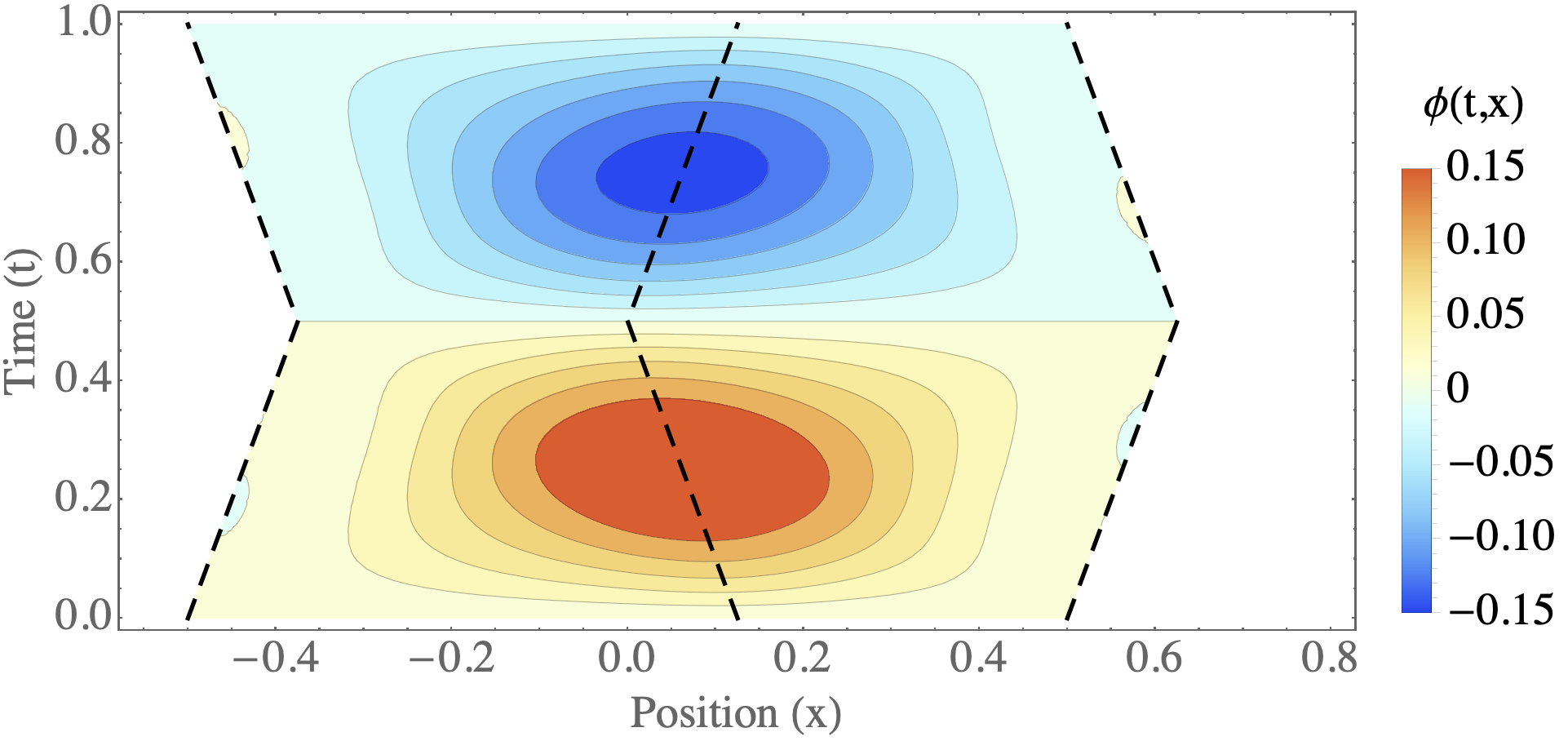}\label{fig:swmap}}
\caption{Spacetime evolution of a swaying oscillon with $v_0=0.25$, reconstructed from a truncated Fourier series containing the first five modes ($n \le 5$). Central zig-zag line denotes position of the maximum $x_0(t)=v_0\frac{T}{2}-y(t)$.}\label{swayingmap}
\end{figure}

In Fig.~\ref{swayingmap}, we plot the gradient map of the Fourier series truncated at $n \le 5$, representing a swaying oscillon with $v_0 = 0.25$. The maximum of the oscillon moves in the opposite direction of its swaying boundaries following the trajectory
\[
x_0(t)=v_0\frac{T}{2}-y(t).
\]
where $y(t)$ was introduced in \eqref{eq:y-trajectory}.

\section{Summary and Remarks}
\label{sec:summary}
In this paper, we have presented a new approach to describing compact oscillons in the signum-Gordon model. By combining Fourier series and Fourier transform methods, we obtain the compact oscillon as an infinite sum of discrete Fourier modes. The key to solving this problem lies in the observation that the signum function characterizing the oscillon solution acts as a spatially localized window function, which periodically switches its sign at the exact instants when the field itself vanishes. Capitalizing on this pre-established signum term, we apply a temporal Fourier series ansatz aligned with the Fourier expansion of the signum source term. This source term selects only odd harmonics; consequently, the resulting Fourier series contains solely odd modes that collectively form the exact oscillon.

Using the Fourier transform, we map the spatial coefficients into the reciprocal space and solve the resulting algebraic equations. Crucially, the direct algebraic solution does not initially constitute a valid distributional solution to the differential equation. To obtain the physically correct solution, we establish a precise rule for treating the poles in the complex plane by defining appropriate contour closures in accordance with the Cauchy residue theorem and Jordan's lemma. The resulting analytical expression vanishes identically outside the compact support, rigorously recovering the compacton boundary.

Interestingly, this general solution allows for configurations representing $N$ side-by-side, localized compact oscillons. For the simplest case, $N=1$, we recover a single compact oscillon. Its spatial profiles on the domain $\vert{}x\vert{} \le T/2$ are given by $1 + \cos(\omega_n x)$ and are modulated by the temporal coefficients $\sin(\omega_n t)$, meaning the oscillon can also be understood as a superposition of standing waves.

We have carefully evaluated the accuracy of the truncated Fourier series against the exact, closed-form solution. Remarkably, the first fundamental mode (with frequency $\omega_1 = \frac{2\pi}{T}$) already provides a reasonably good approximation of the exact profile, and this approximation improves significantly upon the inclusion of the next two odd modes. Finally, an analysis of the power spectrum reveals no frequencies other than the fundamental frequency and its higher harmonics. This spectral purity provides a strong argument to discard the hypothesis that the exact oscillon is a composite bound state of simpler, distinct localized structures -- an idea that had been previously suggested by analyzing the spatial profiles of the exact energy and momentum densities.

Next, we addressed the more complex problem of the Fourier representation of a swaying oscillon. While the presence of a swaying motion initially suggests that the spatial profile functions themselves might depend on time, this possibility was ruled out during our investigation, leading to the correct solution presented in Section \ref{sec:Fourierswaying}. Remarkably, we find that a purely temporal dependence governed by $\sin(\omega_n t)$ coefficients is sufficient to capture this dynamics. The swaying motion of the oscillon is instead generated via a dynamic modulation that strictly requires the inclusion of even modes. Because the signum source term contains only odd modes, these even modes are not directly driven by it; however, their presence is crucial as they break the left-right spatial symmetry of the oscillon, thereby producing the characteristic swaying behavior.

We verified our Fourier representation by directly comparing it with the known closed-form solutions in the literature, which are constructed using parabolic partial solutions. Analyzing the approximations obtained by truncating the Fourier series at consecutive modes, we observed that the first fundamental (odd) mode alone cannot capture the asymmetric profile; it merely generates a spatially shifted, symmetric shape. In contrast, incorporating the first even mode ($n=2$) successfully breaks this symmetry and yields the desired asymmetric shape.

As we examined the contribution of higher modes, we found that the convergence of the Fourier series to the exact oscillon slows down significantly as the boundary velocity $v_0$ approaches unity. In this limit, the oscillon profile becomes increasingly triangular, triggering the classic Gibbs phenomenon associated with approximating sharp, non-differentiable features at the boundaries. Finally, we performed a limit analysis for $v_0 \to 0$, rigorously demonstrating the mathematical consistency of the swaying oscillon solution with the static-boundary case investigated earlier.

One of the principal results of this paper is the explicit demonstration that, unlike the non-convergent perturbative series for oscillons in standard analytic potentials, the corresponding expansion for the signum-Gordon oscillon exhibits exact convergence.

Additionally, the approach presented here offers a powerful alternative framework for analyzing compact oscillons. In particular, for the fundamental oscillon with a static boundary, it bypasses the main drawback of traditional signum-Gordon solutions: the tedious matching of multiple localized solutions across light cones. Although our method involves an infinite series of modes, we demonstrate that truncating the expansion to just a few terms--and in some regimes, even the primary mode alone--yields highly accurate approximations of the exact oscillon profiles.
A detailed understanding of the Fourier spectrum of the exact oscillon provides a valuable framework for analyzing other periodic solutions in the signum-Gordon and related models. In particular, many oscillon-type structures observed in numerical experiments may simply represent excited or bound states of multiple constituent oscillons. Resolving this question requires further investigation, and the Fourier analysis developed here serves as a central tool to help unravel this puzzle.

\section*{Declaration on the Use of Generative Artificial Intelligence}
In compliance with the academic integrity frameworks adopted by CAPES and formally instituted by Portaria CNPq n\textordmasculine~2.664/2026, the authors declare that Generative Artificial Intelligence (AI) was utilized during the drafting and manuscript refinement phase of this work. Specifically, the tool ChatGPT (GPT-4o version, accessed via ChatGPT by OpenAI, current version as of October 2026) was used exclusively for the grammatical and stylistic revision of the text. The authors emphasize that the manuscript was originally produced by them, and the results of the revision via ChatGPT were thoroughly checked, verified, and validated by the authors. The authors assume full and sole responsibility for the accuracy, originality, and scientific integrity of the final content presented in this manuscript.

\begin{acknowledgments}
The authors are indebted to A. Wereszczy\'nski for reading the manuscript and providing many valuable comments.
This study was financed in part by the Coordenação de Aperfeiçoamento de Pessoal de Nível Superior -- Brasil (CAPES) -- Finance Code 001. L.R.L. is supported by the grant $2025/20538-2$, S\~ao Paulo Research Foundation (FAPESP)
\end{acknowledgments}

\bibliography{bibliography}

\end{document}